\documentclass[prx,letterpaper,nobalancelastpage,twocolumn,superscriptaddress,nofootinbib,longbibliography]{revtex4-1}

\usepackage{graphicx}
\usepackage{amsmath}
\usepackage{amssymb}
\usepackage[english]{babel}
\usepackage{color}
\usepackage[version=4]{mhchem}
\usepackage[hidelinks]{hyperref}
\usepackage{dsfont}

\usepackage{soul}
\newcommand{\Caltech}{Division of Physics, Mathematics and Astronomy, California Institute of Technology, Pasadena, CA 91125, USA}
\newcommand{\JPL}{Jet Propulsion Laboratory, California Institute of Technology, Pasadena, CA 91109, USA}

\renewcommand{\cite}[1]{\mbox{\citep{#1}}}

\addto\captionsenglish{}
\addto\captionsenglish{}

\begin{document}

\title{An Atomic Array Optical Clock with Single-Atom Readout}
\author{Ivaylo S. Madjarov}
\affiliation{\Caltech}
\author{Alexandre Cooper}
\affiliation{\Caltech}
\author{Adam L. Shaw}
\affiliation{\Caltech}
\author{Jacob P. Covey}
\affiliation{\Caltech}
\author{Vladimir Schkolnik}
\affiliation{\JPL}
\author{Tai Hyun Yoon}\altaffiliation{Permanent address: Department of Physics, Korea University, Seoul 02841, Republic of Korea}
\affiliation{\Caltech}
\author{Jason R. Williams}
\affiliation{\JPL}
\author{Manuel Endres}\email{mendres@caltech.edu}
\affiliation{\Caltech}

\begin{abstract}
Currently, the most accurate and stable clocks use optical interrogation of either a single ion or an ensemble of neutral atoms confined in an optical lattice. Here, we demonstrate a new optical clock system based on an array of individually trapped neutral atoms with single-atom readout, merging many of the benefits of ion and lattice clocks as well as creating a bridge to recently developed techniques in quantum simulation and computing with neutral atoms. We evaluate single-site resolved frequency shifts and short-term stability via self-comparison. Atom-by-atom feedback control enables direct experimental estimation of laser noise contributions. Results agree well with an \textit{ab initio} Monte Carlo simulation that incorporates finite temperature, projective read-out, laser noise, and feedback dynamics. Our approach, based on a tweezer array, also suppresses interaction shifts while retaining a short dead time, all in a comparatively simple experimental setup suited for transportable operation. These results establish the foundations for a third optical clock platform and provide a novel starting point for entanglement-enhanced metrology, quantum clock networks, and applications in quantum computing and communication with individual neutral atoms that require optical clock state control.
\end{abstract}
\maketitle

\section{Introduction}\vspace{-3mm}
Optical clocks --- based on interrogation of ultra-narrow optical transitions in ions or neutral atoms --- have surpassed traditional microwave clocks in both relative frequency stability and accuracy~\cite{Ludlow2015,McGrew2018,Brewer2019,Oelker2019}. They enable new experiments for geodesy~\cite{Grotti2018a,McGrew2018}, fundamental physics~\cite{Blatt2008,Pruttivarasin2015}, and quantum many-body physics~\cite{Scazza2014}, in addition to a prospective redefinition of the SI second~\cite{McGrew2019}. In parallel, single-atom detection and control techniques have propelled quantum simulation and computing applications based on trapped atomic arrays; in particular, ion traps~\cite{Kim2010}, optical lattices~\cite{Gross2017}, and optical tweezers~\cite{Bernien2017, Lienhard2018}. Integrating such techniques into an optical clock would provide atom-by-atom error evaluation, feedback, and thermometry~\cite{Ovsiannikov2011}; facilitate quantum metrology applications, such as quantum-enhanced clocks~\cite{Gil2014,Braverman2019,Kaubruegger2019,Koczor2019} and clock networks~\cite{Komar2014}; and enable novel quantum computation, simulation, and communication architectures that require optical clock state control combined with single atom trapping~\cite{Daley2008,Pagano2018,Covey2019b}. 

As for current optical clock platforms, ion clocks already incorporate single-particle detection and control~\cite{Huntemann2012}, but they typically operate with only a single ion. Research towards multi-ion clocks is ongoing~\cite{Tan2019}. Conversely, optical lattice clocks (OLCs)~\cite{Ludlow2015,McGrew2018,Oelker2019} interrogate thousands of atoms to improve short-term stability, but single-atom detection and control remains an outstanding challenge. An ideal clock system, in this context, would thus merge the benefits of ion and lattice clocks; namely, a large array of isolated atoms that can be read out and controlled individually.

Here we present a prototype of a new optical clock platform based on an atomic array which naturally incorporates single-atom readout of currently $\approx$40 individually trapped neutral atoms. Specifically, we use a magic wavelength 81-site tweezer array stochastically filled with single strontium-88 ($^{88}$Sr) atoms~\cite{Covey2019a}. Employing a repetitive imaging scheme~\cite{Covey2019a}, we stabilize a local oscillator to the optical clock transition~\cite{Taichenachev2006a,Akatsuka2010} with a low dead time of $\approx$100 ms between clock interrogation blocks. 

We utilize single-site and single-atom resolution to evaluate the in-loop performance of our clock system in terms of stability, local frequency shifts, selected systematic effects, and statistical properties. To this end, we define an error signal for single tweezers which we use to measure site-resolved frequency shifts at otherwise fixed parameters. We also evaluate statistical properties of the in-loop error signal, specifically, the dependence of its variance on atom number and correlations between even and odd sites. 

We further implement a standard interleaved self-comparison technique~\cite{Nicholson2015,Al-Masoudi2015} to evaluate systematic frequency shifts with changing external parameters -- specifically trap depth and wavelength -- and find an operational magic condition~\cite{Brown2017a,Origlia2018,Nemitz2019} where the dependence on trap depth is minimized. We further demonstrate a proof-of-principle for extending such self-comparison techniques to evaluate \textit{single-site-resolved} systematic frequency shifts as a function of a changing external parameter. 

Using self-comparison, we evaluate the fractional short-term instability of our clock system to be $2.5\times 10^{-15}/\sqrt{\tau}$. To compare our experimental results with theory predictions, we develop an {\it ab initio} Monte Carlo (MC) clock simulation~(Appendix~\ref{Sec:MonteCarlo}), which directly incorporates laser noise, projective readout, finite temperature, and feedback dynamics, resulting in higher predictive power compared to traditionally used analytical methods~\cite{Ludlow2015}. Our experimental data agree quantitatively with this simulation, indicating that noise processes are well captured and understood at the level of stability we achieve here. Based on the MC model, we predict a fractional instability of (1.9--2.2)$\times$10$^{-15}/\sqrt{\tau}$ for single clock operation, which would have shorter dead time than that in self-comparison. 

We further demonstrate a direct evaluation of the $1/\sqrt{N_A}$ dependence of clock stability with atom number $N_A$, on top of a laser noise dominated background, through an atom-by-atom system-size-selection technique. This measurement and the MC model strongly indicate that the instability is limited by the frequency noise of our local oscillator. We note that the measured instability is comparable to OLCs using similar transportable laser systems~\cite{Koller2017a}.

\begin{figure*}[t!]
	\centering
	\includegraphics[width=12cm]{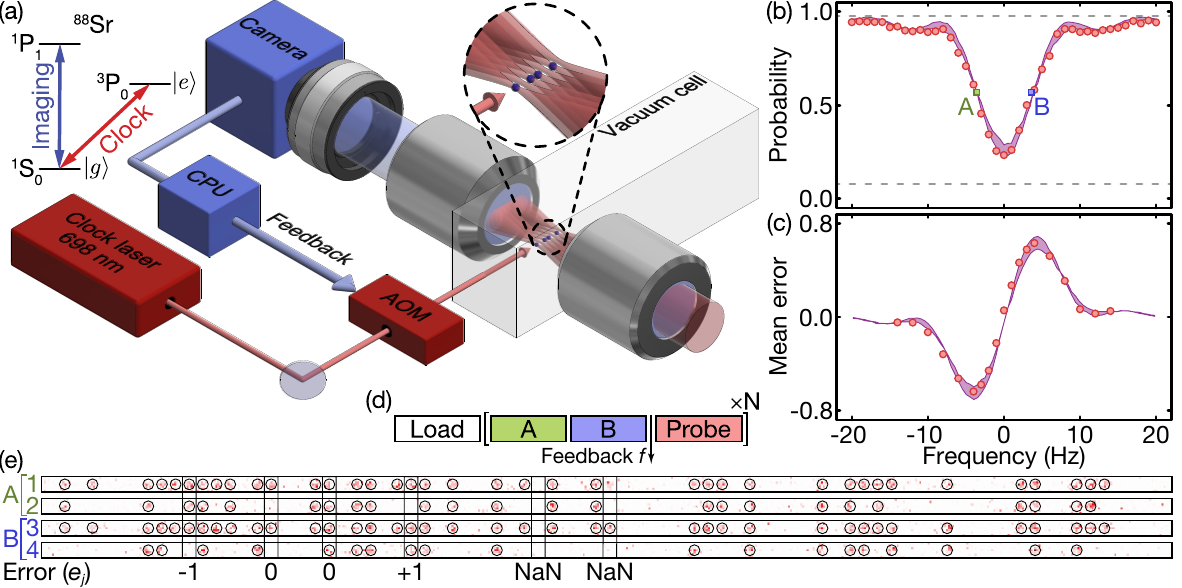}
	\caption{Atomic array optical clock. {(a)}~We interrogate $\approx$40 $^{88}$Sr atoms, trapped in an 81-site tweezer array, on the ultra-narrow clock transition at 698 nm and use high-resolution fluorescence imaging at 461 nm to detect population changes in the clock states (labeled $|g\rangle$ and $|e\rangle$) with single-atom resolution. This information is processed by a central processing unit (CPU) and a feedback signal is applied to the clock laser frequency using an acousto-optic modulator (AOM). {(b)} Tweezer-averaged probability to remain in $|g\rangle$ as a function of frequency offset measured with an in-loop probe sequence (circles). Dashed horizontal lines indicate state-resolved detection fidelities~(Appendix~\ref{Sec:StateDetectionFidelity}). To generate an error signal, we interrogate twice: once below (\textit{A}) and once above (\textit{B}) resonance. {(c)} Tweezer-averaged error signal as a function of frequency offset (circles). The shaded areas in (b) and (c) show results from MC simulations. {(d)}~Simplified experimental sequence, consisting of tweezer loading and $N$-times-repeated \textit{AB} feedback blocks followed by an optional probe block, with $N=10$ throughout. {(e)} To detect the clock state population in block \textit{A}, we take a first image before interrogation to identify which tweezers are occupied and a second image after interrogation to detect which atoms remain in $|g\rangle$ (images 1 and 2). The same procedure is repeated for block \textit{B} (images 3 and 4). We show fluorescence images with identified atoms (circles)~(Appendix~\ref{Sec:Sequence}) and examples of single tweezer error signals $e_j$.}
	\label{FigAtomicArray}
\end{figure*}

We note the very recent, complementary results of Ref.~\cite{Norcia2019} that show seconds-long coherence in a tweezer array filled with $\approx$5 $^{88}$Sr atoms using an ultra-low noise laser without feedback operation. In this and our system, a recently developed repetitive interrogation protocol~\cite{Covey2019a}, similar to that used in ion clocks, provides a short dead time of $\approx$100~ms between interrogation blocks, generally suppressing the impact of laser noise on stability stemming from the Dick effect~\cite{Dick1987}. Utilizing seconds-scale interrogation with such low dead times combined with the feedback operation and realistic upgrade to the system size demonstrated here promises a clock stability that could reach that of state-of-the-art OLCs~\cite{McGrew2018,Oelker2019,Campbell2017,Schioppo2017} in the near-term future, as further discussed in the outlook section.

Concerning systematic effects, the demonstrated atomic array clock has intrinsically suppressed interaction and hopping shifts: First, single atom trapping in tweezers provides immunity to on-site collisions present in one-dimensional OLCs~\cite{Swallows2011}. While three-dimensional OLCs~\cite{Campbell2017} also suppress on-site collisions, our approach retains a short dead time as no evaporative cooling is needed. Further, the adjustable and significantly larger interatomic spacing strongly reduces dipolar interactions~\cite{Chang2004} and hopping effects~\cite{Hutson2019}. We experimentally study effects from tweezer trapping in Sec.~\ref{Sec:SelfCompSyst} and develop a corresponding theoretical model in Appendix~\ref{Sec:LightShift}, but leave a full study of other systematics, not specific to our platform, and a statement of accuracy to future work. In this context, we note that our tweezer system is well-suited for future investigations of black-body radiation shifts via the use of local thermometry with Rydberg states~\cite{Ovsiannikov2011}.

The results presented here and in Ref.~\cite{Norcia2019} provide the foundation for establishing a third optical clock platform promising competitive stability, accuracy, and robustness, while incorporating single-atom detection and control techniques in a natural fashion. We expect this to be a crucial development for applications requiring advanced control and read-out techniques in many-atom quantum systems, as discussed in more detail in the outlook section. 

\vspace{-5mm}\section{Functional principle}\vspace{-3mm}

The basic functional principle is as follows. We generate a tweezer array with linear polarization and $2.5~\mu$m site-to-site spacing in an ultra-high vacuum glass cell using an acousto-optic deflector (AOD) and a high-resolution imaging system~(Fig.~\ref{FigAtomicArray}a)\cite{Covey2019a}. 
The tweezer array wavelength is tuned to a magic trapping configuration close to $813.4$ nm, as described below. 
We load the array from a cold atomic cloud and subsequently induce light-assisted collisions to eliminate higher trap occupancies~\cite{Cooper2018a,Covey2019a}. 
As a result, $\approx$40 of the tweezers are stochastically filled with a single atom. We use a recently demonstrated narrow-line Sisyphus cooling scheme~\cite{Covey2019a} to cool the atoms to an average transverse motional occupation number of $\bar{n}\approx0.66$, measured with clock sideband spectroscopy~(Appendix~\ref{Sec:SidebandThermometry}). The atoms are then interrogated twice on the clock transition, once below (\textit{A}) and once above (\textit{B}) resonance, to obtain an error signal quantifying the frequency offset from the resonance center (Fig.~\ref{FigAtomicArray}b,c). We use this error signal to feedback to a frequency shifter in order to stabilize the frequency of the interrogation laser --- acting as a local oscillator --- to the atomic clock transition.  Since our imaging scheme has a survival fraction of $>$0.998~\cite{Covey2019a}, we perform multiple feedback cycles before reloading the array, each composed of a series of cooling, interrogation, and readout blocks (Fig.~\ref{FigAtomicArray}d). 

For state-resolved readout with single-shot, single-atom resolution, we use a detection scheme composed of two high-resolution images for each of the \textit{A} and \textit{B} interrogation blocks (Fig.~\ref{FigAtomicArray}e)~\cite{Covey2019a}. A first image determines if a tweezer is occupied, followed by clock interrogation. A second image, after interrogation, determines if the atom has remained in the ground state $|g\rangle$. This yields an  instance of an error signal for all tweezers that are occupied at the beginning of both interrogation blocks, while unoccupied tweezers are discounted. For occupied tweezers, we record the $|g\rangle$ occupation numbers $s_{A,j}=\{0,1\}$ and $s_{B,j}=\{0,1\}$ in the images after interrogation with \textit{A} and \textit{B}, respectively, where $j$ is the tweezer index. The difference $e_j=s_{A,j}-s_{B,j}$ defines a single-tweezer error variable taking on three possible values $e_j=\{-1,0,+1\}$ indicating interrogation below, on, or above resonance, respectively. Note that the average of $e_j$ over many interrogations, $\langle e_j\rangle$, is simply an estimator for the difference in transition probability between blocks \textit{A} and \textit{B}.

For feedback to the clock laser, $e_j$ is averaged over all occupied sites in a single \textit{AB} interrogation cycle, yielding an array-averaged error $\bar e = \frac{1}{N_A}\sum_j e_j$, where the sum runs over all occupied tweezers and $N_A$ is the number of present atoms. We add $\bar e$ times a multiplicative factor to the frequency shifter, with the magnitude of this factor optimized to minimize in-loop noise.   

\begin{figure}[t!]
	\centering
	\includegraphics[width=6.0cm]{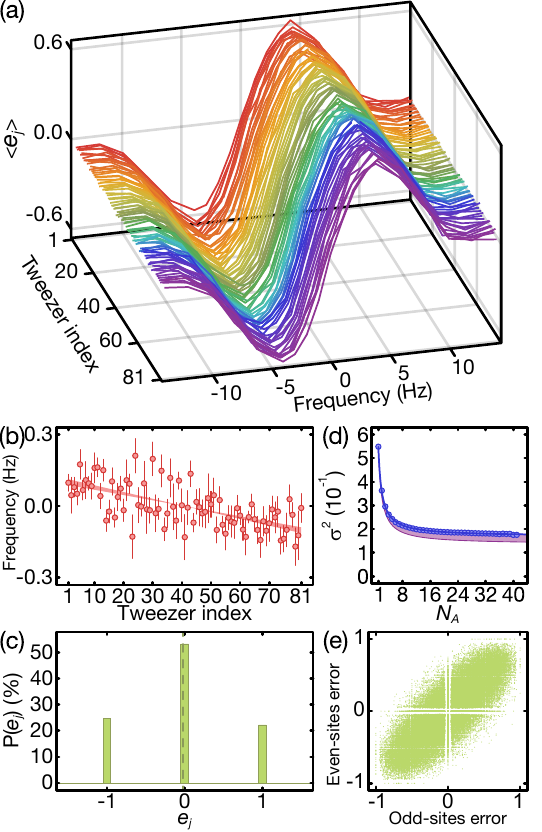}
	\caption{Site-resolved error signal. {(a)}~Repetition-averaged single-tweezer error signal $\langle e_j\rangle$ as a function of frequency offset measured with an in-loop sequence. {(b)}~Fitted zero-crossings as a function of tweezer index for our usual interrogation trap depth of $U_1=245(31) E_r$ where $E_r=h\times3.43$ kHz (circles). Solid lines correspond to theory predictions, with the shaded area resulting from systematic uncertainty in trap depth~(Appendix~\ref{Sec:LightShift}). {(c)}~Ternary probability distribution for $e_j$ for a selected tweezer. The vertical dashed line shows the mean. {(d)} Variance of the error signal as a function of atom number, calculated through post-selection. Solid line is a fit with a $1/N_A$ function plus an offset. Purple region is a MC simulation. {(e)} Plot of correlations between the error signals of even and odd sites.}
	\label{FigErrorSignal}
\end{figure}

\vspace{-3mm}\section{In-loop spectroscopic results}\vspace{-3mm}
We begin by describing results for in-loop detection sequences. Here, feedback is applied to the clock laser (as described before) and probe blocks, for which the interrogation frequency is varied, are added after each feedback cycle.  Using a single probe block with an interrogation time of $110$ ms (corresponding to a $\pi$-pulse on resonance) shows a nearly Fourier-limited line-shape with full-width at half-maximum of $\approx$7 Hz (Fig.~\ref{FigAtomicArray}b). We also use these parameters for the feedback interrogation blocks, with the $\textit{A}$ and $\textit{B}$ interrogation frequencies spaced by a total of $7.6$ Hz. Using the same in-loop detection sequence, we can also directly reveal the shape of the error signal by using two subsequent probe blocks spaced by this frequency difference and scanning a common frequency offset (Fig.~\ref{FigAtomicArray}c). The experimental results are in agreement with MC simulations, which have systematic error denoted as a shaded area throughout, stemming from uncertainty in the noise properties of the interrogation laser~(Appendix~\ref{Sec:FreqNoiseModel}). 

Importantly, these data also exist on the level of individual tweezers, both in terms of averages and statistical fluctuations. As a first example, we show a tweezer-resolved measurement of the repetition-averaged error signal $\langle e_j\rangle$ for all 81 traps (Fig.~\ref{FigErrorSignal}a) as a function of frequency offset. 

Fitting the zero-crossings of $\langle e_j\rangle$ enables us to detect differences in resonance frequency with sub-Hz resolution (Fig.~\ref{FigErrorSignal}b). The results show a small gradient across the array due to the use of an AOD: tweezers are spaced by $500$ kHz in optical frequency, resulting in an approximately linear variation of the clock transition frequency. This effect could be avoided by using a spatial light modulator for tweezer array generation~\cite{Nogrette2014}. We note that the total frequency variation is smaller than the width of our interrogation signal. Such ``sub-bandwidth'' gradients can still lead to noise through stochastic occupation of sites with slightly different frequencies; in our case, we predict an effect at the $10^{-17}$ level. We propose a method to eliminate this type of noise in future clock iterations with a local feedback correction factor in Appendix~\ref{Sec:ErrorCorrection}.

Before moving on, we note that $e_j$ is a random variable with a ternary probability distribution (Fig.~\ref{FigErrorSignal}c) defined for each tweezer. The results in Fig.~\ref{FigErrorSignal}a are the mean of this distribution as a function of frequency offset.  In addition to such averages, having a fully site-resolved signal enables valuable statistical analysis. As an example, we extract the variance of $\bar e$, $\sigma_{\bar e}^2$, for an in-loop probe sequence where the probe blocks are centered around resonance. 

Varying the number of atoms taken into account (via post-selection) shows a $1/N_A$ scaling with a pre-factor dominated by quantum projection noise (QPN)~\cite{Ludlow2015} on top of an offset stemming mainly from laser noise (Fig.~\ref{FigErrorSignal}d). A more detailed analysis reveals that, for our atom number, the relative noise contribution from QPN to $\sigma_{\bar e}$ is only $\approx$26\%~(Appendix~\ref{Sec:StatisticalProperties}). A similar conclusion can be drawn on a qualitative level by evaluating correlations between tweezer resolved errors from odd and even sites, which show a strong common mode contribution indicative of sizable laser noise (Fig.~\ref{FigErrorSignal}e).\vspace{-5mm}

\begin{figure}[t!]
	\centering
	\includegraphics[width=6.0cm]{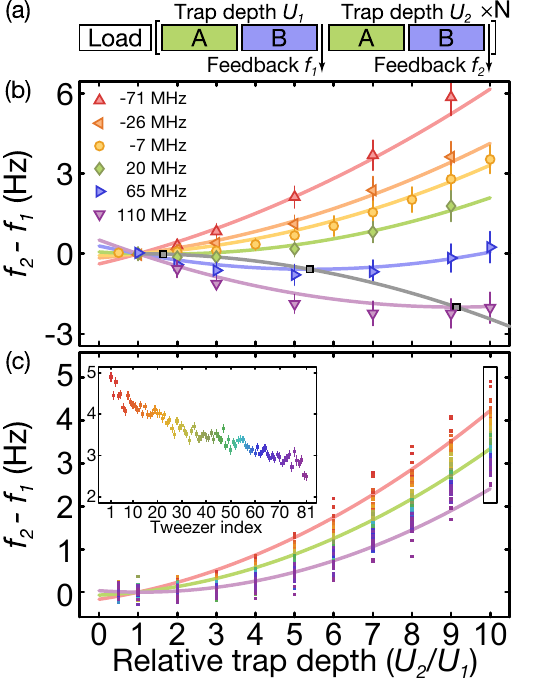}
	\caption{Systematic evaluation of clock shifts with tweezer depth and wavelength. {(a)}~Illustration of interleaved self-comparison, where two independent AOM frequencies ($f_1$ and $f_2$) are updated in an alternating fashion. Respective interrogation blocks are set to two independent tweezer depths $U_1$ and $U_2$. {(b)}~ Average frequency difference $f_2-f_1$ as function of $U_2/U_1$, with $U_1$ fixed to our usual interrogation depth, for multiple frequency offsets of the trapping laser (see legend for color coding). We fit the data with a model for light shifts in optical tweezers (colored lines) with only a single free parameter (for all data simultaneously), accounting for an unknown frequency offset~(Appendix~\ref{Sec:LightShift}). Operational magic intensities are found at the minima of these curves (gray squares and connecting line), which minimize sensitivity to trap depth fluctuations. The trap laser frequency is tuned such that the minimum coincides with our nominal depth. {(c)}~ Combining this technique with the single-tweezer resolved error $\langle e_j\rangle$, we can extract a frequency dependence with trap depth for each tweezer (colored squares).  Solid lines show the expected dependence for the outermost and central tweezers. The data corresponds to the $-7$ MHz set in {(b)}. Inset: Local frequency shifts for $U_2/U_1 = 10$. The color coding of the inset defines the color coding of its containing sub-figure.}
	\label{FigMagic}
\end{figure}

\begin{figure}[t!]
	\centering
	\includegraphics[width=6.0cm]{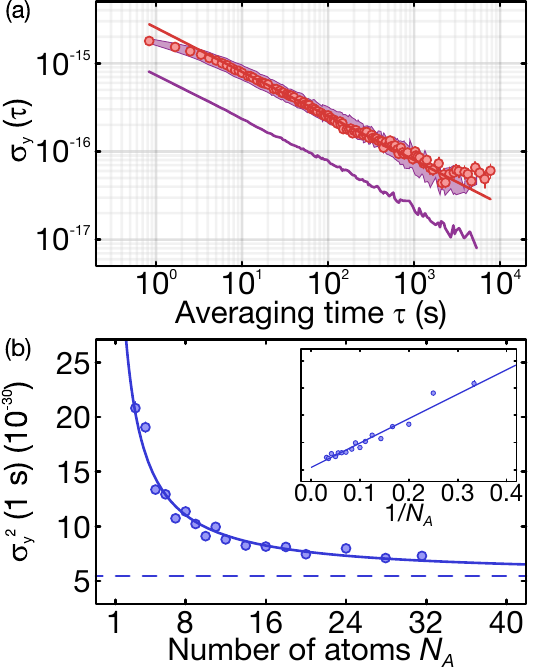}
	\caption{Stability results. {(a)}~ Fractional Allan deviation $\sigma_y$ obtained via self-comparison as a function of integration time $\tau$ (circles). Fitting a $1/\sqrt{\tau}$ behavior past an initial lock onset time (red solid line), we find $2.5\times 10^{-15}/\sqrt{\tau}$. The shaded area denotes MC results. The purple solid line shows the quantum projection noise limit obtained from MC by switching off all other noise sources. {(b)}~ Based on atom-by-atom feedback control, we perform a series of self-comparisons with fixed atom number $N_A$. Shown is the Allan variance $\sigma_y^2$ at one second (from a $1/\sqrt{\tau}$ fit) as a function of $N_A$. Inset: Allan variance $\sigma_y^2$ as a function  of $1/N_A$. Solid lines show a fit with a functional form $\sigma_y^2=\sigma_\infty^2+\sigma_{N_A}^2$, where $\sigma_{N_A}$ scales as $1/\sqrt{N_A}$.}
	\label{FigStability}
\end{figure}

\section{Self-comparison for evaluation of systematic shifts from tweezer trapping}\label{Sec:SelfCompSyst}\vspace{-3mm}
We now turn to an interleaved self-comparison~\cite{Nicholson2015,Al-Masoudi2015}, which we use for stability evaluation and systematic studies. The self-comparison consists of having two feedback loops running in parallel, where feedback is given in an alternating fashion to update two independent AOM frequencies $f_1$ and $f_2$ (Fig.~\ref{FigMagic}a). This is used for a lock-in type evaluation of clock frequency changes with varying parameters. As a specific example, we operate the clock with our usual interrogation trap depth $U_1$ during blocks for feedback to $f_1$ and with a different trap depth $U_2$ during blocks for feedback to $f_2$. The average frequency difference $f_2 - f_1$ now reveals a shift of the clock operation frequency dependent on $U_2$ (Fig.~\ref{FigMagic}b). For optimal clock operation, we find an ``operationally magic'' condition that minimizes sensitivity to trap depth fluctuations~\cite{Brown2017a,Origlia2018,Nemitz2019} by performing two-lock comparisons for different wavelengths (Fig.~\ref{FigMagic}b)~(Appendix~\ref{Sec:LightShift}). We note that this type of standard self-comparison can only reveal array-averaged shifts.

In this context, an important question is how such lock-in techniques can be extended to reveal site-resolved systematic errors as a function of a changing external parameter. To this end we combine the tweezer resolved error signal $\langle e_j\rangle$ with interleaved self-comparison (Fig.~\ref{FigMagic}c). Converting $\langle e_j\rangle$ to frequencies (using measured error functions, such as in Fig.~\ref{FigErrorSignal}a) yields frequency estimators $\delta f_{1,j}$ and $\delta f_{2,j}$ for each tweezer during $f_1$ and $f_2$ feedback blocks, respectively. These estimators correspond to the relative resonance frequency of each tweezer with respect to  the center frequency of the individual locks. Plotting the quantity $\delta f_{2,j}-\delta f_{1,j}+f_2-f_1$ then shows the absolute frequency change of each tweezer as a function of trap depth (Fig.~\ref{FigMagic}c).

\vspace{-5mm}\section{Self-comparison for stability evaluation}\vspace{-3mm}
We use the same self-comparison sequence to evaluate the fractional clock instability by operating both locks with identical conditions (Fig.~\ref{FigStability}a). This approach follows previous clock studies, where true comparison to a second, fully independent clock system was not available~\cite{Nicholson2015,Al-Masoudi2015}. We plot the Allan deviation $\sigma_y$~\cite{Riehle2003} of $y=(f_2-f_1)/(\nu_0\sqrt{2})$ in Fig.~\ref{FigStability}a, where $\nu_0$ is the clock transition frequency and the $\sqrt{2}$ factor is introduced to take into account the addition of noise from two identical sources. The results show a $1/\sqrt{\tau}$ behavior after a lock onset time, where $\tau$ is the averaging time in seconds. Fitting this behavior yields $\sigma_y=2.5\times 10^{-15}/\sqrt{\tau}$, in excellent agreement with MC simulations (Fig.~\ref{FigStability}a). 

Self-comparison evaluates how fast averaging can be performed for systematic studies --- such as the one shown in Fig.~\ref{FigMagic} --- and reveals the impact of various noise sources on short-term stability; however, by design, this technique suppresses slow drifts that are common to the $f_1$ and $f_2$ interrogation blocks. We performed a separate stability analysis by locking $f_1$ to the left half of the array and $f_2$ to the right half of the array \cite{Campbell2017}, a method which is sensitive to slow drifts of gradients, and found no long-term drift of gradients to within our sensitivity~(Appendix~\ref{Sec:SpatiallyResolved}).

Having shown good agreement between our data and MC simulations, we are able to further use the simulation to predict properties of our clock that are not directly experimentally accessible. One of these properties is the true stability of the local oscillator frequency, computed directly by taking the Allan deviation of the simulated laser frequency time traces under feedback. This allows to simulate the stability of single clock operation, which has shorter dead time than the double clock scheme that we use to evaluate stability in experiment. Following this protocol, our simulations predict (1.9--2.2)$\times$10$^{-15}/\sqrt{\tau}$ for the local oscillator stability during single clock operation~(Appendix~\ref{Sec:MonteCarlo}). In this context, we note the results of Ref.~\cite{Norcia2019}, where stability is evaluated by converting a spectroscopic signal into a frequency record (without a closed feedback loop). Based on interrogation with an ultra-low noise laser system, they achieve a short-term stability of 4.7$\times 10^{-16}/\sqrt{\tau}$ with $\approx$5 atoms in tweezers.

Generically, clock stability improves with increasing atom number as $1/\sqrt{N_A}$ through a reduction in readout-noise as long as atoms are uncorrelated. However, in the presence of laser noise --- which is common mode to all atoms --- a limit to stability exists even for an infinite number of atoms~\cite{Ludlow2015}. Intriguingly, we can directly extract such contributions by performing a series of self-comparisons where we adjust the atom number one-by-one (Fig.~\ref{FigStability}b). To this end, we restrict the feedback operation to a subset of atoms in the center of the array with desired size, ignoring the remainder. We are able to achieve stable locking conditions for $N_A\geq3$ with typical feedback parameters. We evaluate the Allan variance at one second as a function of $N_A$ and fit the results with a function $\sigma_y^2=\sigma_\infty^2+\sigma_{N_A}^2$, where $\sigma_{N_A}$ scales as $1/\sqrt{N_A}$. We find $\sigma_{N_A}=6.7\times10^{-15}/\sqrt{N_A\cdot\tau}$ and $\sigma_\infty = 2.3\times10^{-15}/\sqrt{\tau}$, the latter being an estimator for the limit of our clock set by laser noise, in agreement with MC simulation. 
\vspace{-0.5cm}
\section{Outlook}\vspace{-3mm}
Our results merge single-particle readout and control techniques for neutral atom arrays with optical clocks based on ultra-narrow spectroscopy. Such atomic array optical clocks (AOCs) could approach the sub-$10^{-16}/\sqrt{\tau}$ level of stability achieved with OLCs~\cite{McGrew2018,Campbell2017,Schioppo2017,Oelker2019} by increasing interrogation time and atom number. Reaching several hundreds of atoms is realistic with an upgrade to two-dimensional arrays, while Ref.~\cite{Norcia2019} already demonstrated seconds-long interrogation. A further increase in atom number is possible by using a secondary array for readout, created with a non-magic wavelength for which higher power lasers exist~\cite{Norcia2018b, Cooper2018a}. We also envision a system where tweezers are used to ``implant'' atoms, in a structured fashion, into an optical lattice for interrogation and are subsequently used to provide confinement for single-atom readout. Further, the lower dead time of AOCs should help to reduce laser noise contributions to clock stability compared to 3d OLCs~\cite{Campbell2017}, and even zero dead time operation~\cite{Schioppo2017,Campbell2017} in a single machine is conceivable by adding local interrogation. Local interrogation could be achieved through addressing with the main objective or an orthogonal high-resolution path by using spatial-light modulators or acoustic-optic devices. For the case of addressing through the main objective, atoms would likely need to be trapped in an additional lattice to increase longitudinal trapping frequencies.

Concerning systematics, AOCs provide fully site-resolved evaluation combined with an essential mitigation of interaction shifts, while being ready-made for implementing local thermometry using Rydberg states~\cite{Ovsiannikov2011} in order to more precisely determine black-body induced shifts~\cite{Ludlow2015}. In addition, AOCs offer an advanced toolset for generation and detection of entanglement to reach beyond standard quantum limit operation --- either through cavities~\cite{Braverman2019,Norcia2019b} or Rydberg excitation~\cite{Gil2014} --- and for implementing quantum clock networks~\cite{Komar2014}. Further, the demonstrated techniques provide a pathway for quantum computing and communication with neutral alkaline-earth-like atoms~\cite{Daley2008,Scazza2014,Covey2019b}. Finally, features of atomic array clocks, such as experimental simplicity, short dead time, and three-dimensional confinement, make these systems attractive candidates for robust portable clock systems and space-based missions~\cite{Origlia2018}.
\vspace{-3mm}
\section*{Acknowledgments}\label{sec:acknowledgments}
We acknowledge funding provided by the Institute for Quantum Information and Matter, an NSF Physics Frontiers Center (NSF Grant PHY-1733907). This work was supported by the NSF CAREER award (1753386), by the AFOSR YIP (FA9550-19-1-0044), and by the Sloan Foundation. This research was carried out at the Jet Propulsion Laboratory and the California Institute of Technology under a contract with the National Aeronautics and Space Administration and funded through the President's and Director's Research and Development Fund (PDRDF). JPC acknowledges support from the PMA Prize postdoctoral fellowship, AC acknowledges support from the IQIM Postdoctoral Scholar fellowship, and THY acknowledges support from the IQIM Visiting Scholar fellowship and from NRF-2019009974. We acknowledge generous support from Fred Blum.

\setcounter{section}{0}

\twocolumngrid

\renewcommand\appendixname{APPENDIX}
\appendix
\renewcommand\thesection{\Alph{section}}
\renewcommand\thesubsection{\arabic{subsection}}

\section{Monte Carlo simulation}\label{Sec:MonteCarlo}
\subsection{Operation}
We compare the performance of our clock to Monte Carlo (MC) simulations. The simulations include the effects of laser frequency noise, dead time during loading and between interrogations, quantum projection noise, finite temperature, stochastic filling of tweezers, and experimental imperfections such as state-detection infidelity and atom loss. The effects of Raman scattering from the trap and of differential trapping due to hyperpolarizability or trap wavelength shifts from the AOD are not included as they are not expected to be significant at our level of stability.

Rabi interrogation is simulated by time evolving an initial state $|g\rangle$ with the time-dependent Hamiltonian $\hat{H}(t) = \frac{\hbar}{2} (\Omega\sigma_x + (\delta (t) \pm \delta_o) \sigma_z$), where $\Omega$ is the Rabi frequency, $\delta_o$ is an interrogation offset, and $\delta (t)$ is the instantaneous frequency noise defined such that $\delta (t) = \frac{d\phi(t)}{dt}$, where $\phi(t)$ is the optical phase in the rotating frame.  The frequency noise $\delta (t)$ for each Rabi interrogation is sampled from a pre-generated noise trace (Appendix~\ref{Sec:FreqNoiseGeneration}, \ref{Sec:FreqNoiseModel}) with a discrete timestep of $10$ ms. Dead time between interrogations and between array refilling is simulated by sampling from time-separated intervals of this noise trace. Stochastic filling is implemented by sampling the number of atoms $N_A$ from a binomial distribution on each filling cycle, and atom loss is implemented by probabilistically reducing $N_A$ between interrogations.  

To simulate finite temperature, a motional quantum number $n$ is assigned to each of the $N_A$ atoms before each interrogation, where $n$ is sampled from a 1d thermal distribution using our experimentally measured $\bar{n} \approx 0.66$ (Appendix~\ref{Sec:SidebandThermometry}). Here, $n$ represents the motional quantum number along the axis of the interrogating clock beam. For each of the unique values of $n$ that were sampled, a separate Hamiltonian evolution is carried out with a modified Rabi frequency given by $\Omega_n = \Omega e^{-\frac{\eta^2}{2}}L_n(\eta^2)$~\cite{Wineland1979}, where $\eta = \frac{2\pi}{\lambda_{clock}}\sqrt{\frac{\hbar}{2m\omega}}$ is the Lamb-Dicke parameter, $L_n$ is the $n$-th order Laguerre polynomial, and $\Omega$ is the bare Rabi frequency valid in the limit of infinitely tight confinement. 

At the end of each interrogation, excitation probabilities $p_e(n) = |\langle e | \psi_n \rangle|^2$ are computed from the final states for each $n$. State-detection infidelity is simulated by defining adjusted excitation probabilities $\tilde{p}_e(n) \equiv f_e p_e(n) + (1-f_g)(1-p_e(n))$, where $f_g$ and $f_e$ are the ground and excited state detection fidelities (Appendix~\ref{Sec:StateDetectionFidelity}), respectively. To simulate readout of the the $j$-th atom on the $i$-th interrogation, a Bernoulli trial with probability $\tilde{p}_e(n_j)$ is performed, producing a binary readout value $s_{j,i}$. An error signal $\bar{e} = \frac{1}{N_A} \sum_j (s_{j,i-1} - s_{j,i})$ is produced every two interrogation cycles by alternating the sign of $\delta_o$ on alternating interrogation cycles. This error signal produces a control signal (using the same gain factor as used in experiment) which is summed with the generated noise trace for the next interrogation cycle, closing the feedback loop.  

\vspace{-3mm}
\subsection{Generating frequency noise traces}\label{Sec:FreqNoiseGeneration}
Using a model of the power spectral density of our clock laser's frequency noise (Sec~\ref{Sec:FreqNoiseModel}), we generate random frequency noise traces in the time domain~\cite{deLeseleuc2018} for use in the Monte Carlo simulation. Given the power spectral density of frequency noise $S_\nu(f)$, we generate a complex one-sided amplitude spectrum $A_\nu(f) = e^{i\phi(f)}\sqrt{2S_\nu(f)\Delta f}$, where $\phi(f)$ is sampled from a uniform distribution in $[0,2\pi)$ for each $f$ and $\Delta f$ is the frequency discretization. This is converted to a two-sided amplitude spectrum by defining $A_\nu(-f) = A_\nu^*(f)$. Finally, a time trace $\nu(t) = \mathcal{F}\{A(f)\}(t) + \nu_l(t)$ is produced by taking a fast Fourier transform (FFT) of $A(f)$ and adding an experimentally calibrated linear drift term $\nu_l(t)$.

\begin{figure}[t!]
	\centering
	\includegraphics[width=5.5cm]{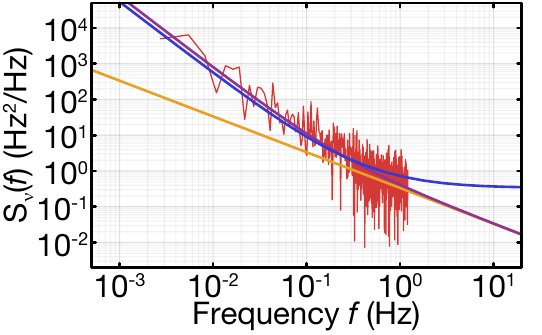}
	\caption{
	Frequency noise spectrum of the clock laser.~
	Power spectral density of the frequency noise of our clock laser measured from a beat signal with a reference laser over a 42-hour period (red trace). Our theoretical estimate of the thermal noise contribution is plotted in yellow. Plotted also are our best- (purple) and worst- (blue) case models for total frequency noise, as used in Monte Carlo simulations.  }
	\label{FigLaserNoise}
\end{figure}

\vspace{-3mm}
\subsection{Frequency noise model}\label{Sec:FreqNoiseModel}
The power spectral density of the frequency noise of our clock laser is modeled by the sum of contributions from random walk frequency modulation (RWFM) noise ($f^{-2}$), flicker frequency modulation (FFM) noise ($f^{-1}$), and white frequency modulation (WFM) noise ($f^{0}$), such that  $S_\nu(f)=\alpha f^{-2}+\beta f^{-1}+\gamma f^{0}$. We obtain these parameters through an estimation of the thermal noise of our reference cavity and a fit of a partially specified frequency noise power spectral density obtained via beating our laser with a reference laser~(Fig.~\ref{FigLaserNoise}). Due to remaining large uncertainty in the white noise floor of our laser, we define a worst- and best-case noise model. The range between these models is the dominant source of uncertainty in our Monte Carlo simulations. 

FFM noise results from thermal mechanical fluctuations of the reference cavity~\cite{Numata2004,Jiang2011}. By estimating the noise contribution from the ultra-low expansion spacer, fused silica mirrors, and their reflective coating, we estimate a fractional frequency instability of $\sigma_y=1.6\times10^{-15}$ at $1$ s, which corresponds to a frequency noise power spectral density of $\beta f^{-1}=0.34$~Hz$^2/$Hz at $f=1$ Hz. 

As a worst case noise model, we assume a cross-over frequency from FFM to WFM noise at $1$ Hz~(Fig.~\ref{FigLaserNoise}), such that $\gamma=\beta f^{-1}=0.34$~Hz$^2/$Hz, and we estimate a frequency noise power spectral density of $\alpha f^{-2}=0.05$~Hz$^2/$Hz at $1$ Hz for RWFM noise. As a best case noise model, assuming no cross-over from FFM to WFM noise (such that $\gamma=0.00$~Hz$^2/$Hz) we estimate a frequency noise power spectral density for RWFM noise of $\alpha f^{-2} = 0.08$~Hz$^2/$Hz at $f = 1$ Hz. We note that the difference in predicted clock stability between the best and worst case model is relatively minor. This indicates that dominant contributions to clock instability stem from frequencies where we have experimental frequency noise data and where both models exhibit similar frequency noise. This is confirmed by an analytical Dick noise analysis~\cite{Dick1987} (not shown).

\section{Experimental details}
\vspace{-3mm}
\subsection{Experimental system} \label{Sec:ExpSystem}
Our strontium apparatus is described in detail in Refs.~\cite{Covey2019a,Cooper2018a}. Strontium-88 atoms from an atomic beam oven are slowed and cooled to a few microkelvin temperature by a 3d magneto-optical trap operating first on the broad dipole-allowed $^1$S$_0\leftrightarrow{}^1$P$_1$ transition at $461$ nm and then on the narrow spin-forbidden $^1$S$_0\leftrightarrow{}^3$P$_1$ transition at $689$ nm. Strontium atoms are filled into a 1d array of 81 optical tweezers at $\lambda_T=813.4$ nm, which is the magic wavelength for the doubly-forbidden $^1$S$_0\leftrightarrow{}^3$P$_0$ optical clock transition. The tweezers have Gaussian waist radii of $800(50)$ nm and an array spacing of $2.5~\mu$m. During filling, cooling, and imaging (state detection), the trap depth is $2447(306)~E_r$. Here $E_r$ is the tweezer photon recoil energy, given by $E_r=h^2/(2m\lambda_T^2)$, where $h$ is Planck's constant and $m$ is the mass of $^{88}$Sr. The tweezer depth is determined from the measured waist and the radial trapping frequency found from sideband measurements on the clock transition (discussed in more detail in Appendix~\ref{Sec:SidebandThermometry}). After parity projection, each tweezer has a 0.5 probability of containing a single atom, or otherwise being empty. Thus, the total number of atoms $N_A$ after each filling cycle of the experiment follows a binomial distribution with mean number of atoms $\bar{N}_A=40.5$.
\vspace{-3mm}
\subsection{Clock laser system}
Our clock laser is based on a modified portable clock laser system (Stable Laser Systems) composed of an external cavity diode laser (Moglabs) stabilized to an isolated, high-finesse optical cavity using the Pound-Drever-Hall scheme and electronic feedback to the laser diode current and piezoelectric transducer. The optical cavity is a 50~mm cubic cavity~\cite{Webster2011} made of ultra-low expansion glass maintained at the zero-crossing temperature of $40.53~^{\circ}$C with mirror substrates made of fused silica with a finesse of $F>300,000$ at $698$ nm. The clock laser light passes through a first AOM in double-pass configuration, injects an anti-reflection coated laser diode (Sacher Lasertechnik GmbH, SAL-0705-020), passes through a second AOM, and goes through a 10~m long fiber to the main experiment with a maximum output optical power of $20$ mW. The first AOM is used for shifting and stabilizing the frequency of the clock laser, whereas the second AOM is used for intensity-noise and fiber-noise cancellation. The clock laser light has a Gaussian waist radius of $600~\mu$m along the tweezer array. This large width is chosen to minimize gradients in clock intensity across the array arising from slight beam angle misalignments.

\begin{figure}[t!]
	\centering
	\includegraphics[width=5.5cm]{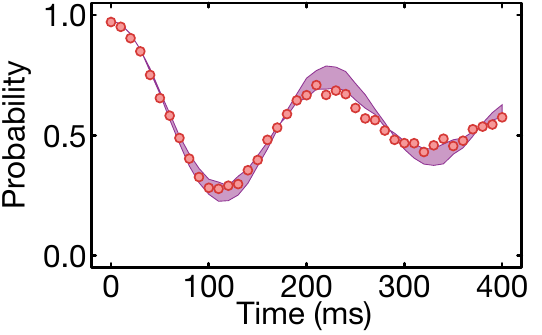}
	\caption{
	Rabi oscillations on the clock transition with $\pi$-pulse length of 110 ms. Each point is probed directly after stabilizing the clock laser with a feedback sequence as described in the main text. The shaded area denotes Monte Carlo results.}
	\label{FigRabi}
\end{figure}

\vspace{-3mm}
\subsection{Bosonic clock transition}\label{Sec:BosonicClock}
Optical excitation of the $^1$S$_0\leftrightarrow{}^3$P$_0$ clock transmission in a bosonic alkaline-earth-like atom is facilitated by applying a bias magnetic field $B$~\cite{Taichenachev2006a}. This field creates a small admixture of $^3$P$_1$ into $^3$P$_0$, and results in a Rabi frequency of $\Omega_R/2\pi=\alpha\sqrt{I}|B|$, where $I$ is the intensity of the clock probe beam and $\alpha$ is the coupling constant. For $^{88}$Sr, $\alpha=198$ Hz/T(mW/cm$^2$)$^{1/2}$~\cite{Taichenachev2006a}. The probe beam induces an AC Stark shift $\Delta\nu_P=kI$, where $k=-18$ mHz/(mW/cm$^2$) for $^{88}$Sr~\cite{Taichenachev2006a}. The magnetic field gives rise to a quadratic Zeeman shift $\Delta\nu_B=\beta B^2$, where $\beta=-23.3$ MHz/T$^2$ for $^{88}$Sr~\cite{Taichenachev2006a}. 

We choose $B\approx900~\mu$T, for which $\Delta\nu_B\approx-18.9$ Hz, and we choose $I\approx1560$ mW/cm$^2$, for which $\Delta\nu_P\approx-28.1$ Hz. The quoted values for $B$ and $I$ are experimentally calibrated by measuring $\Delta\nu_B$ and $\Delta\nu_P$ via two-clock self-comparison~(Sec.~\ref{Sec:SelfCompSyst}) where the value of the systematic parameter in the second rail is varied. We fit the measured frequency shift to a quadratic model for the magnetic shift and to a linear model for the probe shift (not shown) and extrapolate both fits to the known zero values of the systematic parameters, thus extracting $\Delta \nu_B$ and $\Delta \nu_P$. 

We note that our measured $\pi$-time of 110~ms~(Fig.~\ref{FigRabi}) is longer than what would be expected from the calibrated beam intensity. This is likely explained by spectral impurity of the interrogating light, which has servo-induced sidebands at $\approx$600~kHz. These sidebands are spectrally resolved enough so as to not affect clock interrogation, but still contribute to the probe light shift of the transition frequency. 

\vspace{-3mm}
\subsection{Interrogation sequence}\label{Sec:Sequence}
We confirm the presence of atoms in each tweezer using fluorescence imaging for 30 ms on the 461 nm transition while cooling on the 689 nm transition and repumping atoms out of the metastable $^3$P$_{0,2}$ states. This imaging procedure initializes the atoms in the $^1$S$_0$ electronic ground state $|g\rangle$. We then further cool the atoms for 10 ms using attractive Sisyphus cooling~\cite{Covey2019a} on the 689 nm transition and adiabatically ramp down to a trap depth of $245(31)~E_r$ for 4 ms. We apply a weak bias magnetic field of $B\approx900~\mu$T along the transverse direction of the tweezer array to enable direct optical excitation of the doubly-forbidden clock transition at $698$ nm~\cite{Taichenachev2006a,Barber2006}. After interrogating the clock transition for 110 ms (Fig.~\ref{FigRabi}), we adiabatically ramp the trap depth back up to $2447(306)~E_r$ to detect the population of atoms in $|g\rangle$ using fluorescence imaging for 30 ms without repumping on the $^3$P$_0\leftrightarrow{}^3$S$_1$ transition. This interrogation sequence is repeated a number of times before the array is refilled with atoms. 
\vspace{-3mm}
\subsection{Clock state detection fidelity}\label{Sec:StateDetectionFidelity}
Based on the approach demonstrated in Ref.~\cite{Covey2019a},
we analyze the fidelity of detecting atoms in the $^1$S$_0$ ($|g\rangle$) and $^3$P$_0$ ($|e\rangle$) states under these imaging conditions. We diagnose our state-detection fidelity with two consecutive images. In the first image, we detect atoms in $|g\rangle$ by turning off the $^3$P$_0\leftrightarrow{}^3$S$_1$ repump laser such that atoms in $|e\rangle$ in principle remain in $|e\rangle$ and do not scatter photons~\cite{Covey2019a}.
Hence, if we find a signal in the first image, we identify the state as $|g\rangle$. In the second image, we turn the $^3$P$_0\leftrightarrow{}^3$S$_1$ repump laser back on to detect atoms in both $|g\rangle$ and $|e\rangle$. Thus, if an atom is not detected in the first image but appears in the second image we can identify it as $|e\rangle$. If neither of the images shows a signal we identify the state as ``no-atom''.

The inaccuracy of this scheme is dominated by off-resonant scattering of the tweezer light when atoms are shelved in $|e\rangle$ during the first image. By pumping atoms into $|e\rangle$ before imaging, we observe that they decay back to $|g\rangle$ with a time constant of $\tau_p=370(4)$ ms at our imaging trap depth of $2447(306)~E_r$. This leads to events in the first image where $|e\rangle$ atoms are misidentified as $|g\rangle$ atoms. Additionally, atoms in $|g\rangle$ can be misidentified as $|e\rangle$ if they are pumped to $|e\rangle$ in the first image. We measure this misidentification probability by initializing atoms in $|g\rangle$ and counting how often we identify them as $|e\rangle$. Using this method, we place a lower bound for the probability of correctly identifying $|e\rangle$ as $f_e \equiv e^{-t/\tau_p}=0.922(1)$ and we directly measure the probability of correctly identifying $|g\rangle$ as $f_g = 0.977(2)$. These values are shown in Fig.~\ref{FigAtomicArray}b as dashed lines.
\vspace{-3mm}
\subsection{Stabilization to the atomic signal}
The clock laser is actively stabilized to the atomic signal using a digital control system. The frequency deviation of the clock laser from the atomic transition is estimated from a two-point measurement of the Rabi spectroscopy signal at $\delta_o/2\pi=\pm3.8$ Hz for an interrogation time of $110$ ms, which produces an experimentally measured lineshape with a full-width at half-maximum of $7$ Hz. $\bar{e}$ is converted into a frequency correction by multiplying it by a factor of $\kappa=3$~Hz. We choose $\kappa$ to be the largest value possible before the variance of the error signal in an in-loop probe sequence begins to grow. Feedback is performed by adding the frequency correction to the frequency of the RF synthesizer (Moglabs ARF421) driving the first AOM along the clock beam path.

\begin{figure}[t!]
	\centering
	\includegraphics[width=5.5cm]{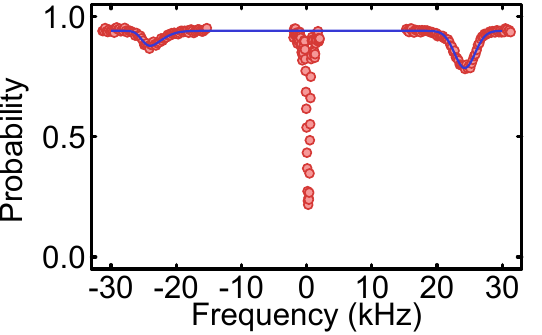}
	\caption{
	Clock sideband thermometry.~
	Array-averaged radial sideband spectrum of the optical clock transition, taken with a carrier Rabi frequency of $\approx$360~Hz. A narrow carrier stands in between two broader sidebands on the red and blue detuned sides. Sideband broadening is due mainly to small inhomogeneities in the array. A suppressed red sideband indicates significant motional ground state population. The solid line is a simultaneous fit to two skewed Gaussians. From the ratio of the area under the red sideband to that under the blue sideband, we obtain $\bar{n}\approx 0.66$. The carrier is probed for an interrogation time of 1.4~ms while the sidebands are probed for 3.3~ms. }
	\label{FigSideband}
\end{figure}

\vspace{-3mm}
\subsection{Sideband thermometry on the clock transition}\label{Sec:SidebandThermometry}
We perform sideband thermometry on the clock transition (Fig.~\ref{FigSideband}) using the same beam used to interrogate the atoms for clock operation. Using a standard technique of taking the ratio of the integrated area under the first red and blue sidebands \cite{Han2018}, we obtain $\bar{n} \approx 0.66$ along the direction of the interrogation beam, oriented along one of the tight radial axes of our tweezers. From the sideband separation, we measure a trap frequency of $\omega \approx 2\pi\times 24.5$ kHz. These values are measured after cooling on the narrow $^1$S$_0 \leftrightarrow{}^3$P$_1$ transition for $10$ ms~\cite{Covey2019a} in a trap of depth $2447(306)~E_r$ and adiabatically ramping down to our clock interrogation depth of $245(31)~E_r$.

We note that the clock transition is sufficiently narrow to observe sub-kHz inhomogeneities of trap frequencies between tweezers. This precision afforded by the clock transition allows for detailed knowledge about inhomogeneities in the array, and we envision using it for fine corrections and uniformization of an array in the future. However, for the purpose of thermometry, we broaden the clock line to a degree that these inhomogeneities are unresolved on an array-averaged level so we may obtain a spectrum that can be easily fit and integrated. Specifically, we use a much higher magnetic field of $\approx$75~mT to obtain a carrier Rabi frequency $\approx$360~Hz at the same optical intensity.
\vspace{-3mm}
\subsection{Evaluating Allan deviations}
Repeated interrogation introduces a bimodal distribution in the time between feedback events due to the periodic refilling of the array. To account for this variation, we approximate that all feedbacks are equally spaced in time with $\Delta t\approx835$ ms. This introduces a slight error $\Delta\tau\approx100$ ms for all $\tau$, though this error is inconsequential for fitting the long time Allan deviation behavior. We fit all Allan deviations from $\tau=10$ s to $\tau=100$ s, using $\sigma_y=A/\sqrt{\tau}$, with free parameter $A=\sigma_y(\tau=1$ s).

\section{Statistical properties of the error signal} \label{Sec:StatisticalProperties}
\subsection{Probability distribution function}
In the absence of additional noise and given $N_A$ atoms, the probability of finding $N_g$ atoms in the ground state after a single clock interrogation block is given by the binomial distribution $P_\mathcal{B}(N_g;N_A,p)$, where $p$ is the probability of detecting an atom in its ground state following clock interrogation. The probability of measuring a given error signal $\bar{e}=\Delta N_g/N_A$ is thus given by the probability of measuring the difference atom number $\Delta N_g=N_g^A-N_g^B$, where $N_g^A$ ($N_g^B$) is the number of atoms detected in the ground state after the \textit{A} (\textit{B}) interrogation blocks. It can be shown that the probability distribution for $\Delta N_g$ is given by the convolution of two binomial distributions, $P_{\ast}(\Delta N_g;N_A,p_A,p_B)=\sum_{N}P_\mathcal{B}(N;N_A,p_B)P_\mathcal{B}(N-\Delta N_g;N_A,p_A)$. This discrete distribution has support on $\{-N_A, -N_A+1,\cdots,N_A\}$ with $2N_A+1$ non-zero values. Thus, the probability distribution for $\bar{e}$ is given by $P(\bar{e}; N_A,p_A,p_B)=P_{\ast}(\bar{e}N_A;N_A,p_A,p_B)$. In the absence of statistical correlation between the \textit{A} and \textit{B} interrogation blocks, this distribution has a mean $\mu_{\bar{e}}=(p_B-p_A)$ and a variance $\sigma^2_{\bar{e}}=(p_A (1-p_A) + p_B(1-p_B))/N_A$.
\vspace{-3mm}
\subsection{Additional noise}
In the presence of noise, such as laser noise or finite temperature, the excitation probability $p_A$ and $p_B$ fluctuates from repetition to repetition. These fluctuations can be accounted for by introducing a joint probability density function $\pi(p_A,p_B)$, so that 
\begin{equation}
\begin{split}
P(\bar{e};N_A)=&\int dp_A dp_B \bigg(\pi(p_A,p_B)\\
&\times P(\bar{e};N_A,p_A,p_B)\bigg)\\~=&\langle P(\bar{e};N_A,p_A,p_B)\rangle,
\end{split}
\end{equation}
where $\langle\cdot\rangle$ denotes statistical averaging over $\pi(p_A,p_B)$. 
Assuming the mean of $P(\bar e,N_A)$ to be zero, which is equivalent to $\langle p_A\rangle=\langle p_B\rangle\equiv\langle p\rangle$, and the variance of $p_A$ and $p_B$ to be equal, $\sigma_{p_A}^2=\sigma_{p_B}^2\equiv\sigma_{p}^2$, it can be shown that the variance of $P(\bar e,N_A)$ is given by
\begin{equation}
\sigma_{\bar e}^2=2(\langle p\rangle(1-\langle p\rangle)-\sigma_{p}^2)/N_A + 2(\sigma_{p}^2- C),
\end{equation}
where $C$ is a correlation function between $p_A$ and $p_B$ defined as $C=\langle p_A p_B\rangle- \langle p_A\rangle \langle p_B\rangle$.
\vspace{-3mm}
\subsection{Experimental data}
We can directly extract the correlation function $C$ through the results of images (2) and (4) for valid tweezers (Fig.~\ref{FigAtomicArray}e). We explicitly confirm that $C$ is independent of the number of atoms used per \textit{AB} interrogation cycle and extract $C=-0.025$. The anti-correlation is an indication of laser noise. Note that, in contrast to $C$, $\sigma_{p}^2$ is not directly experimentally accessible as it is masked by QPN. The fit to the variance of the error signal (Fig.~\ref{FigErrorSignal}d) yields $\sigma_{\bar e}^2=0.379/N_A+0.169$. We can thus use the fitted offset of $0.169$ combined with the knowledge of $C$ to extract $\sigma_{p}^2=0.059$. We can alternatively use the fitted coefficient of the $1/N_A$ term of $0.379$ with the measured $\langle p\rangle=0.41$ to extract $\sigma_{p}^2=0.052$. To determine the contribution from QPN versus other noise sources in the standard deviation of the error signal, we take $\sigma_{\bar{e},QPN}=\sqrt{2\langle p\rangle(1-\langle p\rangle)/N_A}/\sigma_{\bar{e}},$ which for $N_A=40.5~$yields $\sigma_{\bar{e},QPN}=0.26,$ as quoted in the main text.

\section{Exploiting single-site resolved signals}
\subsection{Atom number dependent stability}
To study the performance of our clock as a function of atom number, we can choose to use only part of our full array for clock operation (Fig.~\ref{FigStability}b). We preferentially choose atoms near the center of the array to minimize errors due to gradients in the array e.g. from the AOD. Due to the stochastic nature of array filling, we generally use different tweezers during each filling cycle such as to always compute a signal from a fixed number of atoms. When we target a large number of atoms, some repetitions have an atom number lower than the target due to the stochastic nature of array filling, resulting in a mean atom number slightly smaller than the target as well as a small fluctuation in atom number. The data points in Fig.~\ref{FigStability}b show the mean atom numbers used for clock operation, with error bars around these means (denoting the standard deviation of atom number) being smaller than the marker size.

\vspace{-3mm}
\subsection{Clock comparison between two halves of the array}\label{Sec:SpatiallyResolved}
\begin{figure}[t!]
	\centering
	\includegraphics[width=5.5cm]{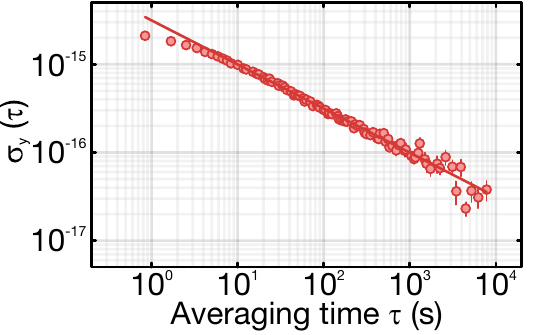}
	\caption{
	Spatially-resolved clock comparison.~
	The fractional Allan deviation from an asynchronous clock comparison between the left and right half of our array. Fitting a $1/\sqrt{\tau}$ behavior past an initial lock onset time, we find $3.1\times 10^{-15}/\sqrt{\tau}$, slightly higher than the result measured for a self-comparison of the full array (Fig.~\ref{FigStability}). Importantly, we see no upturn for times approaching $10^4$~s and below the $10^{-16}$ level, indicating that slowly-varying drifts of gradients across the array do not contribute to instability up to our sensitivity. }
	\label{FigSpatiallyResolved}
\end{figure}

We use the ability to lock to a subset of occupied traps to perform stability analysis that is sensitive to slow drifts of gradients across the array (such as from external fields or spatial variations in trap homogeneity). In this case, we lock $f_1$ to traps 1-40 and lock $f_2$ to traps 42-81, such that noise sources which vary across the array will show a divergence in the Allan deviation at long enough times. As shown in Fig.~\ref{FigSpatiallyResolved}, we perform this analysis for times approaching $\tau = 10^4$~s and down to the $\sigma_y=1\times10^{-16}$ level, and observe no violation of the $\sigma_y\propto1/\sqrt{\tau}$ behavior. Thus, we conclude that such temporal variations in gradients are not a resolvable systematic for our current experiment. However, this analysis will prove useful when using an upgraded system for which stability at the $\sigma_y=10^{-17}$ level or lower becomes problematic. In principle, the lock could be done on a single trap position at a time, which would allow trap-by-trap systematics to be analyzed.

\vspace{-3mm}
\subsection{\textit{In situ} error correction}\label{Sec:ErrorCorrection}
Single-site resolution offers the opportunity both to analyze single-atom signals, as discussed in the main text, and to modify such signals before using them for feedback. As an example, the AOD introduces a spatial gradient in trap frequencies across the array, leading to a spatial variation in zero-crossings of the error signal (as shown in Fig.~\ref{FigErrorSignal}b) and subsequently leading to an increase in the Allan deviation at the $\sigma_y\approx10^{-17}$ level due to stochastic trap loading. While this effect is not currently significant in our experiment, it and other array inhomogeneities may be visible to future experiments with increased stability.

Therefore, we propose that this problem can be corrected (for inhomogeneities within the probe bandwidth) by adjusting the error signal $e_j$ of each tweezer by a correction factor \textit{before} calculating the array-averaged $\bar{e}$ that will produce feedback for the local oscillator. For instance, consider the modification $\bar{e}_f = \frac{1}{N_A}\sum_j \zeta_j e_j-f_{0,j}$, where $\bar{e}_f$ is the tweezer-averaged error in Hz, $\zeta_j$ is a tweezer-resolved conversion factor such as could be obtained from Fig.~\ref{FigErrorSignal}a, and $f_{0,j}$ is the tweezer-resolved zero-crossing of the error signal. This new formulation mitigates inhomogeneity without any physical change to the array. While physically enforcing array uniformity is ideal, this is a tool that can simplify the complexity of correcting experimental systematics.

\section{Tweezer-induced light shifts}\label{Sec:LightShift}
\vspace{-1mm}
Several previous studies have analyzed the polarizability and hyperpolarizability of alkaline-earth-like atoms, including $^{88}$Sr, in magic wavelength optical lattices~\cite{Brown2017a,Origlia2018,Nemitz2019,Katori2015}.
In their analyses, these studies include the effect of finite atom temperature by Taylor expanding the lattice potential in powers of $\sqrt{I}$ ($I$ is the lattice intensity) in the vicinity of the magic wavelength~\cite{Katori2015}. We repeat this derivation for an optical tweezer instead of an optical lattice. 

The Gaussian tweezer intensity (assumed to have azimuthal symmetry) is given by $I(\rho,z)=I_0(w_0/w(z))^2e^{-2\rho^2/w(z)^2}$, where $w_0$ is the beam waist, $I_0=2P_0/\pi w_0^2$ is the maximum intensity, $P_0$ is the beam power, $w(z)=w_0\sqrt{1+(z/z_R)^2}$, and $z_R=\pi w_0^2/\lambda_T$ is the Rayleigh range. The trapping potential is determined from this intensity $I(\rho,z)$ by the electric dipole polarizability $\alpha^{E1}$, the electric quadrupole and magnetic dipole polarizabilities $\alpha^{qm}=\alpha^{E2}+\alpha^{M1}$, and the hyperpolarizability effect $\beta I^2$.

By considering a harmonic approximation in the $x$- and $y$-directions as well as harmonic and anharmonic terms in the $z$-direction, we arrive at the following expression for the differential light shift of the clock transition in an optical tweezer, where $\rho = \sqrt{x^2+y^2}$ and $n_\rho (= n_x + n_y)$ and $n_z$ are vibrational quantum number along the radial and axial directions, respectively:
\begin{eqnarray}
&&h\nu_{LS}=-\bigg[\bigg(\frac{\partial}{\partial\nu}\tilde{\alpha}^{E1}\bigg)\delta_{L}+\left(\frac{w_0}{z_{R}}\right)^2\bigg(n_\rho+\frac{1}{2}\bigg)^2 \tilde{\beta}\nonumber\\&&+\sqrt{2}\bigg(\frac{w_0}{z_{R}}\bigg)^3\bigg(n_z+\frac{1}{2}\bigg)\bigg(n_\rho+\frac{1}{2}\bigg)\tilde{\beta}\nonumber \\&&+\frac{3}{8}\bigg(\frac{w_0}{z_{R}}\bigg)^4\bigg(n_z^2+n_z+\frac{1}{2}\bigg)\tilde{\beta}\ \nonumber\bigg]u\
\nonumber \\&&+\bigg[2\sqrt{2}\bigg(\frac{w_0}{z_{R}}\bigg)\bigg(n_\rho+\frac{1}{2}\bigg)+\bigg(\frac{w_0}{z_{R}}\bigg)^2\bigg(n_z+\frac{1}{2}\bigg)\bigg]\ \nonumber \\&& \times\bigg[\bigg(\frac{\partial}{\partial\nu}\tilde{\alpha}^{E1}\bigg)\delta_{L}+\tilde{\alpha}^{{qm}}\bigg]u^{1/2}\
\nonumber \\&&+\bigg[2\sqrt{2}\bigg(\frac{w_0}{z_{R}}\bigg)\bigg(n_\rho+\frac{1}{2}\bigg)+\bigg(\frac{w_0}{z_{R}}\bigg)^2\bigg(n_z+\frac{1}{2}\bigg)\bigg]\tilde{\beta}u^{3/2}\nonumber \\
&&-\tilde{\beta}u^2,
\label{eq:EqLightShift}
\end{eqnarray}
where $\tilde{\alpha}^{E1}=\Delta\alpha^{E1}(E_R/\alpha^{E1})$, $\Delta\alpha^{E1}=\alpha^{E1}_e-\alpha^{E1}_g$ is the differential $E1$ polarizability; $\tilde{\alpha}^{qm}=\Delta\alpha^{qm}(E_R/\alpha^{E1})$, where $\Delta\alpha^{qm}$ is the differential $E2$ and $M1$ polarizability; $\tilde{\beta}=\Delta\beta(E_R/\alpha^{E1})^2$, where $\Delta\beta$ is the differential hyperpolarizability; $u=I/(E_R/\alpha^{E1})$ is the tweezer depth.

We use this formula to predict the light shifts studied in the main text (Fig.~\ref{FigMagic}). As we find the results to be mostly insensitive to temperature for low temperatures, we assume zero temperature for simplicity. We allow a single fit parameter, which is an overall frequency shift due to uncertainty in the optical frequency of the trapping light. The other factors are taken from previous studies, as summarized in Table~\ref{table:TableLightShift}. 

\begin{table*}[h!]
\caption{Light shifts of a $^{88}$Sr clock. The fits and predictions based on Eq.~\ref{eq:EqLightShift} use the following values from previous studies. }
\centering
\begin{tabular}{p{5cm}{c}{c}{c}{c}}
\hline\hline
Quantity&Symbol&Unit&Value&Reference\\
\hline
Magic trapping frequency& $\nu_T$ & MHz & 368\,554\,732(11) & \cite{Origlia2018} \\
Hyperpolarizability difference& $\frac{1}{h}\tilde{\beta}$ & $\mu$Hz & 0.45(10) & \cite{LeTargat2013} \\
Slope of $\tilde{\alpha}^{E1}$ & $\frac{1}{h}\frac{\partial\tilde{\alpha}^{E1}}{\partial\nu}$ & ~ & 19.3$\times$10$^{-12}$ & \cite{Origlia2018} \\
Electric dipole polarizability & $\tilde{\alpha}^{E1}$ & kHz/(kW/cm$^2$) & 46.5976(13) & \cite{Middelmann2012} \\
Differential electric quadrupole and magnetic dipole polarizabilities & $\frac{1}{h}\tilde{\alpha}^{qm}$ & mHz & 0.0(3) & \cite{Westergaard2011} \\
\hline\hline
\end{tabular}
\label{table:TableLightShift}
\end{table*}


\clearpage


\begin{thebibliography}{56}%
\makeatletter
\providecommand \@ifxundefined [1]{%
 \@ifx{#1\undefined}
}%
\providecommand \@ifnum [1]{%
 \ifnum #1\expandafter \@firstoftwo
 \else \expandafter \@secondoftwo
 \fi
}%
\providecommand \@ifx [1]{%
 \ifx #1\expandafter \@firstoftwo
 \else \expandafter \@secondoftwo
 \fi
}%
\providecommand \natexlab [1]{#1}%
\providecommand \enquote  [1]{``#1''}%
\providecommand \bibnamefont  [1]{#1}%
\providecommand \bibfnamefont [1]{#1}%
\providecommand \citenamefont [1]{#1}%
\providecommand \href@noop [0]{\@secondoftwo}%
\providecommand \href [0]{\begingroup \@sanitize@url \@href}%
\providecommand \@href[1]{\@@startlink{#1}\@@href}%
\providecommand \@@href[1]{\endgroup#1\@@endlink}%
\providecommand \@sanitize@url [0]{\catcode `\\12\catcode `\$12\catcode
  `\&12\catcode `\#12\catcode `\^12\catcode `\_12\catcode `\%12\relax}%
\providecommand \@@startlink[1]{}%
\providecommand \@@endlink[0]{}%
\providecommand \url  [0]{\begingroup\@sanitize@url \@url }%
\providecommand \@url [1]{\endgroup\@href {#1}{\urlprefix }}%
\providecommand \urlprefix  [0]{URL }%
\providecommand \Eprint [0]{\href }%
\providecommand \doibase [0]{http://dx.doi.org/}%
\providecommand \selectlanguage [0]{\@gobble}%
\providecommand \bibinfo  [0]{\@secondoftwo}%
\providecommand \bibfield  [0]{\@secondoftwo}%
\providecommand \translation [1]{[#1]}%
\providecommand \BibitemOpen [0]{}%
\providecommand \bibitemStop [0]{}%
\providecommand \bibitemNoStop [0]{.\EOS\space}%
\providecommand \EOS [0]{\spacefactor3000\relax}%
\providecommand \BibitemShut  [1]{\csname bibitem#1\endcsname}%
\let\auto@bib@innerbib\@empty
\bibitem [{\citenamefont {Ludlow}\ \emph {et~al.}(2015)\citenamefont {Ludlow},
  \citenamefont {Boyd}, \citenamefont {Ye}, \citenamefont {Peik},\ and\
  \citenamefont {Schmidt}}]{Ludlow2015}%
  \BibitemOpen
  \bibfield  {author} {\bibinfo {author} {\bibfnamefont {Andrew~D.}\
  \bibnamefont {Ludlow}}, \bibinfo {author} {\bibfnamefont {Martin~M.}\
  \bibnamefont {Boyd}}, \bibinfo {author} {\bibfnamefont {Jun}\ \bibnamefont
  {Ye}}, \bibinfo {author} {\bibfnamefont {E.}~\bibnamefont {Peik}}, \ and\
  \bibinfo {author} {\bibfnamefont {P.~O.}\ \bibnamefont {Schmidt}},\
  }\bibfield  {title} {\enquote {\bibinfo {title} {{Optical atomic clocks}},}\
  }\href {\doibase 10.1103/RevModPhys.87.637} {\bibfield  {journal} {\bibinfo
  {journal} {Rev. Mod. Phys.}\ }\textbf {\bibinfo {volume} {87}},\ \bibinfo
  {pages} {637--701} (\bibinfo {year} {2015})}\BibitemShut {NoStop}%
\bibitem [{\citenamefont {McGrew}\ \emph {et~al.}(2018)\citenamefont {McGrew},
  \citenamefont {Zhang}, \citenamefont {Fasano}, \citenamefont
  {Sch{\"{a}}ffer}, \citenamefont {Beloy}, \citenamefont {Nicolodi},
  \citenamefont {Brown}, \citenamefont {Hinkley}, \citenamefont {Milani},
  \citenamefont {Schioppo}, \citenamefont {Yoon},\ and\ \citenamefont
  {Ludlow}}]{McGrew2018}%
  \BibitemOpen
  \bibfield  {author} {\bibinfo {author} {\bibfnamefont {W.~F.}\ \bibnamefont
  {McGrew}}, \bibinfo {author} {\bibfnamefont {X.}~\bibnamefont {Zhang}},
  \bibinfo {author} {\bibfnamefont {R.~J.}\ \bibnamefont {Fasano}}, \bibinfo
  {author} {\bibfnamefont {S.~A.}\ \bibnamefont {Sch{\"{a}}ffer}}, \bibinfo
  {author} {\bibfnamefont {K.}~\bibnamefont {Beloy}}, \bibinfo {author}
  {\bibfnamefont {D.}~\bibnamefont {Nicolodi}}, \bibinfo {author}
  {\bibfnamefont {R.~C.}\ \bibnamefont {Brown}}, \bibinfo {author}
  {\bibfnamefont {N.}~\bibnamefont {Hinkley}}, \bibinfo {author} {\bibfnamefont
  {G.}~\bibnamefont {Milani}}, \bibinfo {author} {\bibfnamefont
  {M.}~\bibnamefont {Schioppo}}, \bibinfo {author} {\bibfnamefont {T.~H.}\
  \bibnamefont {Yoon}}, \ and\ \bibinfo {author} {\bibfnamefont {A.~D.}\
  \bibnamefont {Ludlow}},\ }\bibfield  {title} {\enquote {\bibinfo {title}
  {{Atomic clock performance enabling geodesy below the centimetre level}},}\
  }\href {\doibase 10.1038/s41586-018-0738-2} {\bibfield  {journal} {\bibinfo
  {journal} {Nature}\ }\textbf {\bibinfo {volume} {564}},\ \bibinfo {pages}
  {87--90} (\bibinfo {year} {2018})}\BibitemShut {NoStop}%
\bibitem [{\citenamefont {Brewer}\ \emph {et~al.}(2019)\citenamefont {Brewer},
  \citenamefont {Chen}, \citenamefont {Hankin}, \citenamefont {Clements},
  \citenamefont {Chou}, \citenamefont {Wineland}, \citenamefont {Hume},\ and\
  \citenamefont {Leibrandt}}]{Brewer2019}%
  \BibitemOpen
  \bibfield  {author} {\bibinfo {author} {\bibfnamefont {S.~M.}\ \bibnamefont
  {Brewer}}, \bibinfo {author} {\bibfnamefont {J.-S.}\ \bibnamefont {Chen}},
  \bibinfo {author} {\bibfnamefont {A.~M.}\ \bibnamefont {Hankin}}, \bibinfo
  {author} {\bibfnamefont {E.~R.}\ \bibnamefont {Clements}}, \bibinfo {author}
  {\bibfnamefont {C.~W.}\ \bibnamefont {Chou}}, \bibinfo {author}
  {\bibfnamefont {D.~J.}\ \bibnamefont {Wineland}}, \bibinfo {author}
  {\bibfnamefont {D.~B.}\ \bibnamefont {Hume}}, \ and\ \bibinfo {author}
  {\bibfnamefont {D.~R.}\ \bibnamefont {Leibrandt}},\ }\bibfield  {title}
  {\enquote {\bibinfo {title} {{$^{27}$Al$^+$ Quantum-Logic Clock with a
  Systematic Uncertainty below $10^{-18}$}},}\ }\href {\doibase
  10.1103/PhysRevLett.123.033201} {\bibfield  {journal} {\bibinfo  {journal}
  {Phys. Rev. Lett.}\ }\textbf {\bibinfo {volume} {123}},\ \bibinfo {pages}
  {033201} (\bibinfo {year} {2019})}\BibitemShut {NoStop}%
\bibitem [{\citenamefont {Oelker}\ \emph {et~al.}(2019)\citenamefont {Oelker},
  \citenamefont {Hutson}, \citenamefont {Kennedy}, \citenamefont {Sonderhouse},
  \citenamefont {Bothwell}, \citenamefont {Goban}, \citenamefont {Kedar},
  \citenamefont {Sanner}, \citenamefont {Robinson}, \citenamefont {Marti},
  \citenamefont {Matei}, \citenamefont {Legero}, \citenamefont {Giunta},
  \citenamefont {Holzwarth}, \citenamefont {Riehle}, \citenamefont {Sterr},\
  and\ \citenamefont {Ye}}]{Oelker2019}%
  \BibitemOpen
  \bibfield  {author} {\bibinfo {author} {\bibfnamefont {E.}~\bibnamefont
  {Oelker}}, \bibinfo {author} {\bibfnamefont {R.~B.}\ \bibnamefont {Hutson}},
  \bibinfo {author} {\bibfnamefont {C.~J.}\ \bibnamefont {Kennedy}}, \bibinfo
  {author} {\bibfnamefont {L.}~\bibnamefont {Sonderhouse}}, \bibinfo {author}
  {\bibfnamefont {T.}~\bibnamefont {Bothwell}}, \bibinfo {author}
  {\bibfnamefont {A.}~\bibnamefont {Goban}}, \bibinfo {author} {\bibfnamefont
  {D.}~\bibnamefont {Kedar}}, \bibinfo {author} {\bibfnamefont
  {C.}~\bibnamefont {Sanner}}, \bibinfo {author} {\bibfnamefont {J.~M.}\
  \bibnamefont {Robinson}}, \bibinfo {author} {\bibfnamefont {G.~E.}\
  \bibnamefont {Marti}}, \bibinfo {author} {\bibfnamefont {D.~G.}\ \bibnamefont
  {Matei}}, \bibinfo {author} {\bibfnamefont {T.}~\bibnamefont {Legero}},
  \bibinfo {author} {\bibfnamefont {M.}~\bibnamefont {Giunta}}, \bibinfo
  {author} {\bibfnamefont {R.}~\bibnamefont {Holzwarth}}, \bibinfo {author}
  {\bibfnamefont {F.}~\bibnamefont {Riehle}}, \bibinfo {author} {\bibfnamefont
  {U.}~\bibnamefont {Sterr}}, \ and\ \bibinfo {author} {\bibfnamefont
  {J.}~\bibnamefont {Ye}},\ }\bibfield  {title} {\enquote {\bibinfo {title}
  {{Demonstration of $4.8 \times 10^{-17}$ stability at 1 s for two independent
  optical clocks}},}\ }\href {\doibase 10.1038/s41566-019-0493-4} {\bibfield
  {journal} {\bibinfo  {journal} {Nat. Photonics}\ } (\bibinfo {year} {2019}),\
  10.1038/s41566-019-0493-4}\BibitemShut {NoStop}%
\bibitem [{\citenamefont {Grotti}\ \emph {et~al.}(2018)\citenamefont {Grotti},
  \citenamefont {Koller}, \citenamefont {Vogt}, \citenamefont {H{\"{a}}fner},
  \citenamefont {Sterr}, \citenamefont {Lisdat}, \citenamefont {Denker},
  \citenamefont {Voigt}, \citenamefont {Timmen}, \citenamefont {Rolland},
  \citenamefont {Baynes}, \citenamefont {Margolis}, \citenamefont {Zampaolo},
  \citenamefont {Thoumany}, \citenamefont {Pizzocaro}, \citenamefont {Rauf},
  \citenamefont {Bregolin}, \citenamefont {Tampellini}, \citenamefont
  {Barbieri}, \citenamefont {Zucco}, \citenamefont {Costanzo}, \citenamefont
  {Clivati}, \citenamefont {Levi},\ and\ \citenamefont
  {Calonico}}]{Grotti2018a}%
  \BibitemOpen
  \bibfield  {author} {\bibinfo {author} {\bibfnamefont {Jacopo}\ \bibnamefont
  {Grotti}}, \bibinfo {author} {\bibfnamefont {Silvio}\ \bibnamefont {Koller}},
  \bibinfo {author} {\bibfnamefont {Stefan}\ \bibnamefont {Vogt}}, \bibinfo
  {author} {\bibfnamefont {Sebastian}\ \bibnamefont {H{\"{a}}fner}}, \bibinfo
  {author} {\bibfnamefont {Uwe}\ \bibnamefont {Sterr}}, \bibinfo {author}
  {\bibfnamefont {Christian}\ \bibnamefont {Lisdat}}, \bibinfo {author}
  {\bibfnamefont {Heiner}\ \bibnamefont {Denker}}, \bibinfo {author}
  {\bibfnamefont {Christian}\ \bibnamefont {Voigt}}, \bibinfo {author}
  {\bibfnamefont {Ludger}\ \bibnamefont {Timmen}}, \bibinfo {author}
  {\bibfnamefont {Antoine}\ \bibnamefont {Rolland}}, \bibinfo {author}
  {\bibfnamefont {Fred~N.}\ \bibnamefont {Baynes}}, \bibinfo {author}
  {\bibfnamefont {Helen~S.}\ \bibnamefont {Margolis}}, \bibinfo {author}
  {\bibfnamefont {Michel}\ \bibnamefont {Zampaolo}}, \bibinfo {author}
  {\bibfnamefont {Pierre}\ \bibnamefont {Thoumany}}, \bibinfo {author}
  {\bibfnamefont {Marco}\ \bibnamefont {Pizzocaro}}, \bibinfo {author}
  {\bibfnamefont {Benjamin}\ \bibnamefont {Rauf}}, \bibinfo {author}
  {\bibfnamefont {Filippo}\ \bibnamefont {Bregolin}}, \bibinfo {author}
  {\bibfnamefont {Anna}\ \bibnamefont {Tampellini}}, \bibinfo {author}
  {\bibfnamefont {Piero}\ \bibnamefont {Barbieri}}, \bibinfo {author}
  {\bibfnamefont {Massimo}\ \bibnamefont {Zucco}}, \bibinfo {author}
  {\bibfnamefont {Giovanni~A.}\ \bibnamefont {Costanzo}}, \bibinfo {author}
  {\bibfnamefont {Cecilia}\ \bibnamefont {Clivati}}, \bibinfo {author}
  {\bibfnamefont {Filippo}\ \bibnamefont {Levi}}, \ and\ \bibinfo {author}
  {\bibfnamefont {Davide}\ \bibnamefont {Calonico}},\ }\bibfield  {title}
  {\enquote {\bibinfo {title} {{Geodesy and metrology with a transportable
  optical clock}},}\ }\href {\doibase 10.1038/s41567-017-0042-3} {\bibfield
  {journal} {\bibinfo  {journal} {Nat. Phys.}\ }\textbf {\bibinfo {volume}
  {14}},\ \bibinfo {pages} {437--441} (\bibinfo {year} {2018})}\BibitemShut
  {NoStop}%
\bibitem [{\citenamefont {Blatt}\ \emph {et~al.}(2008)\citenamefont {Blatt},
  \citenamefont {Ludlow}, \citenamefont {Campbell}, \citenamefont {Thomsen},
  \citenamefont {Zelevinsky}, \citenamefont {Boyd}, \citenamefont {Ye},
  \citenamefont {Baillard}, \citenamefont {Fouch{\'{e}}}, \citenamefont {{Le
  Targat}}, \citenamefont {Brusch}, \citenamefont {Lemonde}, \citenamefont
  {Takamoto}, \citenamefont {Hong}, \citenamefont {Katori},\ and\ \citenamefont
  {Flambaum}}]{Blatt2008}%
  \BibitemOpen
  \bibfield  {author} {\bibinfo {author} {\bibfnamefont {S.}~\bibnamefont
  {Blatt}}, \bibinfo {author} {\bibfnamefont {A.~D.}\ \bibnamefont {Ludlow}},
  \bibinfo {author} {\bibfnamefont {G.~K.}\ \bibnamefont {Campbell}}, \bibinfo
  {author} {\bibfnamefont {J.~W.}\ \bibnamefont {Thomsen}}, \bibinfo {author}
  {\bibfnamefont {T.}~\bibnamefont {Zelevinsky}}, \bibinfo {author}
  {\bibfnamefont {M.~M.}\ \bibnamefont {Boyd}}, \bibinfo {author}
  {\bibfnamefont {J.}~\bibnamefont {Ye}}, \bibinfo {author} {\bibfnamefont
  {X.}~\bibnamefont {Baillard}}, \bibinfo {author} {\bibfnamefont
  {M.}~\bibnamefont {Fouch{\'{e}}}}, \bibinfo {author} {\bibfnamefont
  {R.}~\bibnamefont {{Le Targat}}}, \bibinfo {author} {\bibfnamefont
  {A.}~\bibnamefont {Brusch}}, \bibinfo {author} {\bibfnamefont
  {P.}~\bibnamefont {Lemonde}}, \bibinfo {author} {\bibfnamefont
  {M.}~\bibnamefont {Takamoto}}, \bibinfo {author} {\bibfnamefont {F.-L.}\
  \bibnamefont {Hong}}, \bibinfo {author} {\bibfnamefont {H.}~\bibnamefont
  {Katori}}, \ and\ \bibinfo {author} {\bibfnamefont {V.~V.}\ \bibnamefont
  {Flambaum}},\ }\bibfield  {title} {\enquote {\bibinfo {title} {{New Limits on
  Coupling of Fundamental Constants to Gravity Using $^{87}$Sr Optical Lattice
  Clocks}},}\ }\href {\doibase 10.1103/PhysRevLett.100.140801} {\bibfield
  {journal} {\bibinfo  {journal} {Phys. Rev. Lett.}\ }\textbf {\bibinfo
  {volume} {100}},\ \bibinfo {pages} {140801} (\bibinfo {year}
  {2008})}\BibitemShut {NoStop}%
\bibitem [{\citenamefont {Pruttivarasin}\ \emph {et~al.}(2015)\citenamefont
  {Pruttivarasin}, \citenamefont {Ramm}, \citenamefont {Porsev}, \citenamefont
  {Tupitsyn}, \citenamefont {Safronova}, \citenamefont {Hohensee},\ and\
  \citenamefont {H{\"{a}}ffner}}]{Pruttivarasin2015}%
  \BibitemOpen
  \bibfield  {author} {\bibinfo {author} {\bibfnamefont {T.}~\bibnamefont
  {Pruttivarasin}}, \bibinfo {author} {\bibfnamefont {M.}~\bibnamefont {Ramm}},
  \bibinfo {author} {\bibfnamefont {S.~G.}\ \bibnamefont {Porsev}}, \bibinfo
  {author} {\bibfnamefont {I.~I.}\ \bibnamefont {Tupitsyn}}, \bibinfo {author}
  {\bibfnamefont {M.~S.}\ \bibnamefont {Safronova}}, \bibinfo {author}
  {\bibfnamefont {M.~A.}\ \bibnamefont {Hohensee}}, \ and\ \bibinfo {author}
  {\bibfnamefont {H.}~\bibnamefont {H{\"{a}}ffner}},\ }\bibfield  {title}
  {\enquote {\bibinfo {title} {{Michelson-Morley analogue for electrons using
  trapped ions to test Lorentz symmetry}},}\ }\href {\doibase
  10.1038/nature14091} {\bibfield  {journal} {\bibinfo  {journal} {Nature}\
  }\textbf {\bibinfo {volume} {517}},\ \bibinfo {pages} {592--595} (\bibinfo
  {year} {2015})}\BibitemShut {NoStop}%
\bibitem [{\citenamefont {Scazza}\ \emph {et~al.}(2014)\citenamefont {Scazza},
  \citenamefont {Hofrichter}, \citenamefont {H{\"{o}}fer}, \citenamefont {{De
  Groot}}, \citenamefont {Bloch},\ and\ \citenamefont
  {F{\"{o}}lling}}]{Scazza2014}%
  \BibitemOpen
  \bibfield  {author} {\bibinfo {author} {\bibfnamefont {F.}~\bibnamefont
  {Scazza}}, \bibinfo {author} {\bibfnamefont {C.}~\bibnamefont {Hofrichter}},
  \bibinfo {author} {\bibfnamefont {M.}~\bibnamefont {H{\"{o}}fer}}, \bibinfo
  {author} {\bibfnamefont {P.~C.}\ \bibnamefont {{De Groot}}}, \bibinfo
  {author} {\bibfnamefont {I.}~\bibnamefont {Bloch}}, \ and\ \bibinfo {author}
  {\bibfnamefont {S.}~\bibnamefont {F{\"{o}}lling}},\ }\bibfield  {title}
  {\enquote {\bibinfo {title} {{Observation of two-orbital spin-exchange
  interactions with ultracold SU(N)-symmetric fermions}},}\ }\href {\doibase
  10.1038/nphys3061} {\bibfield  {journal} {\bibinfo  {journal} {Nat. Phys.}\
  }\textbf {\bibinfo {volume} {10}},\ \bibinfo {pages} {779--784} (\bibinfo
  {year} {2014})}\BibitemShut {NoStop}%
\bibitem [{\citenamefont {McGrew}\ \emph {et~al.}(2019)\citenamefont {McGrew},
  \citenamefont {Zhang}, \citenamefont {Leopardi}, \citenamefont {Fasano},
  \citenamefont {Nicolodi}, \citenamefont {Beloy}, \citenamefont {Yao},
  \citenamefont {Sherman}, \citenamefont {Sch{\"{a}}ffer}, \citenamefont
  {Savory}, \citenamefont {Brown}, \citenamefont {R{\"{o}}misch}, \citenamefont
  {Oates}, \citenamefont {Parker}, \citenamefont {Fortier},\ and\ \citenamefont
  {Ludlow}}]{McGrew2019}%
  \BibitemOpen
  \bibfield  {author} {\bibinfo {author} {\bibfnamefont {W.~F.}\ \bibnamefont
  {McGrew}}, \bibinfo {author} {\bibfnamefont {X.}~\bibnamefont {Zhang}},
  \bibinfo {author} {\bibfnamefont {H.}~\bibnamefont {Leopardi}}, \bibinfo
  {author} {\bibfnamefont {R.~J.}\ \bibnamefont {Fasano}}, \bibinfo {author}
  {\bibfnamefont {D.}~\bibnamefont {Nicolodi}}, \bibinfo {author}
  {\bibfnamefont {K.}~\bibnamefont {Beloy}}, \bibinfo {author} {\bibfnamefont
  {J.}~\bibnamefont {Yao}}, \bibinfo {author} {\bibfnamefont {J.~A.}\
  \bibnamefont {Sherman}}, \bibinfo {author} {\bibfnamefont {S.~A.}\
  \bibnamefont {Sch{\"{a}}ffer}}, \bibinfo {author} {\bibfnamefont
  {J.}~\bibnamefont {Savory}}, \bibinfo {author} {\bibfnamefont {R.~C.}\
  \bibnamefont {Brown}}, \bibinfo {author} {\bibfnamefont {S.}~\bibnamefont
  {R{\"{o}}misch}}, \bibinfo {author} {\bibfnamefont {C.~W.}\ \bibnamefont
  {Oates}}, \bibinfo {author} {\bibfnamefont {T.~E.}\ \bibnamefont {Parker}},
  \bibinfo {author} {\bibfnamefont {T.~M.}\ \bibnamefont {Fortier}}, \ and\
  \bibinfo {author} {\bibfnamefont {A.~D.}\ \bibnamefont {Ludlow}},\ }\bibfield
   {title} {\enquote {\bibinfo {title} {{Towards the optical second: verifying
  optical clocks at the SI limit}},}\ }\href {\doibase 10.1364/OPTICA.6.000448}
  {\bibfield  {journal} {\bibinfo  {journal} {Optica}\ }\textbf {\bibinfo
  {volume} {6}},\ \bibinfo {pages} {448} (\bibinfo {year} {2019})}\BibitemShut
  {NoStop}%
\bibitem [{\citenamefont {Kim}\ \emph {et~al.}(2010)\citenamefont {Kim},
  \citenamefont {Chang}, \citenamefont {Korenblit}, \citenamefont {Islam},
  \citenamefont {Edwards}, \citenamefont {Freericks}, \citenamefont {Lin},
  \citenamefont {Duan},\ and\ \citenamefont {Monroe}}]{Kim2010}%
  \BibitemOpen
  \bibfield  {author} {\bibinfo {author} {\bibfnamefont {K.}~\bibnamefont
  {Kim}}, \bibinfo {author} {\bibfnamefont {M.-S.}\ \bibnamefont {Chang}},
  \bibinfo {author} {\bibfnamefont {S.}~\bibnamefont {Korenblit}}, \bibinfo
  {author} {\bibfnamefont {R.}~\bibnamefont {Islam}}, \bibinfo {author}
  {\bibfnamefont {E.~E.}\ \bibnamefont {Edwards}}, \bibinfo {author}
  {\bibfnamefont {J.~K.}\ \bibnamefont {Freericks}}, \bibinfo {author}
  {\bibfnamefont {G.-D.}\ \bibnamefont {Lin}}, \bibinfo {author} {\bibfnamefont
  {L.-M.}\ \bibnamefont {Duan}}, \ and\ \bibinfo {author} {\bibfnamefont
  {C.}~\bibnamefont {Monroe}},\ }\bibfield  {title} {\enquote {\bibinfo {title}
  {{Quantum simulation of frustrated Ising spins with trapped ions}},}\ }\href
  {\doibase 10.1038/nature09071} {\bibfield  {journal} {\bibinfo  {journal}
  {Nature}\ }\textbf {\bibinfo {volume} {465}},\ \bibinfo {pages} {590--593}
  (\bibinfo {year} {2010})}\BibitemShut {NoStop}%
\bibitem [{\citenamefont {Gross}\ and\ \citenamefont
  {Bloch}(2017)}]{Gross2017}%
  \BibitemOpen
  \bibfield  {author} {\bibinfo {author} {\bibfnamefont {Christian}\
  \bibnamefont {Gross}}\ and\ \bibinfo {author} {\bibfnamefont {Immanuel}\
  \bibnamefont {Bloch}},\ }\bibfield  {title} {\enquote {\bibinfo {title}
  {{Quantum simulations with ultracold atoms in optical lattices}},}\ }\href
  {\doibase 10.1126/science.aal3837} {\bibfield  {journal} {\bibinfo  {journal}
  {Science}\ }\textbf {\bibinfo {volume} {357}},\ \bibinfo {pages} {995--1001}
  (\bibinfo {year} {2017})}\BibitemShut {NoStop}%
\bibitem [{\citenamefont {Bernien}\ \emph {et~al.}(2017)\citenamefont
  {Bernien}, \citenamefont {Schwartz}, \citenamefont {Keesling}, \citenamefont
  {Levine}, \citenamefont {Omran}, \citenamefont {Pichler}, \citenamefont
  {Choi}, \citenamefont {Zibrov}, \citenamefont {Endres}, \citenamefont
  {Greiner}, \citenamefont {Vuleti{\'{c}}},\ and\ \citenamefont
  {Lukin}}]{Bernien2017}%
  \BibitemOpen
  \bibfield  {author} {\bibinfo {author} {\bibfnamefont {Hannes}\ \bibnamefont
  {Bernien}}, \bibinfo {author} {\bibfnamefont {Sylvain}\ \bibnamefont
  {Schwartz}}, \bibinfo {author} {\bibfnamefont {Alexander}\ \bibnamefont
  {Keesling}}, \bibinfo {author} {\bibfnamefont {Harry}\ \bibnamefont
  {Levine}}, \bibinfo {author} {\bibfnamefont {Ahmed}\ \bibnamefont {Omran}},
  \bibinfo {author} {\bibfnamefont {Hannes}\ \bibnamefont {Pichler}}, \bibinfo
  {author} {\bibfnamefont {Soonwon}\ \bibnamefont {Choi}}, \bibinfo {author}
  {\bibfnamefont {Alexander~S.}\ \bibnamefont {Zibrov}}, \bibinfo {author}
  {\bibfnamefont {Manuel}\ \bibnamefont {Endres}}, \bibinfo {author}
  {\bibfnamefont {Markus}\ \bibnamefont {Greiner}}, \bibinfo {author}
  {\bibfnamefont {Vladan}\ \bibnamefont {Vuleti{\'{c}}}}, \ and\ \bibinfo
  {author} {\bibfnamefont {Mikhail~D.}\ \bibnamefont {Lukin}},\ }\bibfield
  {title} {\enquote {\bibinfo {title} {{Probing many-body dynamics on a 51-atom
  quantum simulator}},}\ }\href {\doibase 10.1038/nature24622} {\bibfield
  {journal} {\bibinfo  {journal} {Nature}\ }\textbf {\bibinfo {volume} {551}},\
  \bibinfo {pages} {579--584} (\bibinfo {year} {2017})}\BibitemShut {NoStop}%
\bibitem [{\citenamefont {Lienhard}\ \emph {et~al.}(2018)\citenamefont
  {Lienhard}, \citenamefont {de~L{\'{e}}s{\'{e}}leuc}, \citenamefont {Barredo},
  \citenamefont {Lahaye}, \citenamefont {Browaeys}, \citenamefont {Schuler},
  \citenamefont {Henry},\ and\ \citenamefont {L{\"{a}}uchli}}]{Lienhard2018}%
  \BibitemOpen
  \bibfield  {author} {\bibinfo {author} {\bibfnamefont {Vincent}\ \bibnamefont
  {Lienhard}}, \bibinfo {author} {\bibfnamefont {Sylvain}\ \bibnamefont
  {de~L{\'{e}}s{\'{e}}leuc}}, \bibinfo {author} {\bibfnamefont {Daniel}\
  \bibnamefont {Barredo}}, \bibinfo {author} {\bibfnamefont {Thierry}\
  \bibnamefont {Lahaye}}, \bibinfo {author} {\bibfnamefont {Antoine}\
  \bibnamefont {Browaeys}}, \bibinfo {author} {\bibfnamefont {Michael}\
  \bibnamefont {Schuler}}, \bibinfo {author} {\bibfnamefont {Louis-Paul}\
  \bibnamefont {Henry}}, \ and\ \bibinfo {author} {\bibfnamefont {Andreas~M.}\
  \bibnamefont {L{\"{a}}uchli}},\ }\bibfield  {title} {\enquote {\bibinfo
  {title} {{Observing the Space- and Time-Dependent Growth of Correlations in
  Dynamically Tuned Synthetic Ising Models with Antiferromagnetic
  Interactions}},}\ }\href {\doibase 10.1103/PhysRevX.8.021070} {\bibfield
  {journal} {\bibinfo  {journal} {Phys. Rev. X}\ }\textbf {\bibinfo {volume}
  {8}},\ \bibinfo {pages} {021070} (\bibinfo {year} {2018})}\BibitemShut
  {NoStop}%
\bibitem [{\citenamefont {Ovsiannikov}\ \emph {et~al.}(2011)\citenamefont
  {Ovsiannikov}, \citenamefont {Derevianko},\ and\ \citenamefont
  {Gibble}}]{Ovsiannikov2011}%
  \BibitemOpen
  \bibfield  {author} {\bibinfo {author} {\bibfnamefont {Vitali~D.}\
  \bibnamefont {Ovsiannikov}}, \bibinfo {author} {\bibfnamefont {Andrei}\
  \bibnamefont {Derevianko}}, \ and\ \bibinfo {author} {\bibfnamefont {Kurt}\
  \bibnamefont {Gibble}},\ }\bibfield  {title} {\enquote {\bibinfo {title}
  {{Rydberg Spectroscopy in an Optical Lattice: Blackbody Thermometry for
  Atomic Clocks}},}\ }\href {\doibase 10.1103/PhysRevLett.107.093003}
  {\bibfield  {journal} {\bibinfo  {journal} {Phys. Rev. Lett.}\ }\textbf
  {\bibinfo {volume} {107}},\ \bibinfo {pages} {093003} (\bibinfo {year}
  {2011})}\BibitemShut {NoStop}%
\bibitem [{\citenamefont {Gil}\ \emph {et~al.}(2014)\citenamefont {Gil},
  \citenamefont {Mukherjee}, \citenamefont {Bridge}, \citenamefont {Jones},\
  and\ \citenamefont {Pohl}}]{Gil2014}%
  \BibitemOpen
  \bibfield  {author} {\bibinfo {author} {\bibfnamefont {L.~I.~R.}\
  \bibnamefont {Gil}}, \bibinfo {author} {\bibfnamefont {R.}~\bibnamefont
  {Mukherjee}}, \bibinfo {author} {\bibfnamefont {E.~M.}\ \bibnamefont
  {Bridge}}, \bibinfo {author} {\bibfnamefont {M.~P.~A.}\ \bibnamefont
  {Jones}}, \ and\ \bibinfo {author} {\bibfnamefont {T.}~\bibnamefont {Pohl}},\
  }\bibfield  {title} {\enquote {\bibinfo {title} {{Spin Squeezing in a Rydberg
  Lattice Clock}},}\ }\href {\doibase 10.1103/PhysRevLett.112.103601}
  {\bibfield  {journal} {\bibinfo  {journal} {Phys. Rev. Lett.}\ }\textbf
  {\bibinfo {volume} {112}},\ \bibinfo {pages} {103601} (\bibinfo {year}
  {2014})}\BibitemShut {NoStop}%
\bibitem [{\citenamefont {Braverman}\ \emph {et~al.}(2019)\citenamefont
  {Braverman}, \citenamefont {Kawasaki}, \citenamefont
  {Pedrozo-Pe{\~{n}}afiel}, \citenamefont {Colombo}, \citenamefont {Shu},
  \citenamefont {Li}, \citenamefont {Mendez}, \citenamefont {Yamoah},
  \citenamefont {Salvi}, \citenamefont {Akamatsu}, \citenamefont {Xiao},\ and\
  \citenamefont {Vuleti{\'{c}}}}]{Braverman2019}%
  \BibitemOpen
  \bibfield  {author} {\bibinfo {author} {\bibfnamefont {Boris}\ \bibnamefont
  {Braverman}}, \bibinfo {author} {\bibfnamefont {Akio}\ \bibnamefont
  {Kawasaki}}, \bibinfo {author} {\bibfnamefont {Edwin}\ \bibnamefont
  {Pedrozo-Pe{\~{n}}afiel}}, \bibinfo {author} {\bibfnamefont {Simone}\
  \bibnamefont {Colombo}}, \bibinfo {author} {\bibfnamefont {Chi}\ \bibnamefont
  {Shu}}, \bibinfo {author} {\bibfnamefont {Zeyang}\ \bibnamefont {Li}},
  \bibinfo {author} {\bibfnamefont {Enrique}\ \bibnamefont {Mendez}}, \bibinfo
  {author} {\bibfnamefont {Megan}\ \bibnamefont {Yamoah}}, \bibinfo {author}
  {\bibfnamefont {Leonardo}\ \bibnamefont {Salvi}}, \bibinfo {author}
  {\bibfnamefont {Daisuke}\ \bibnamefont {Akamatsu}}, \bibinfo {author}
  {\bibfnamefont {Yanhong}\ \bibnamefont {Xiao}}, \ and\ \bibinfo {author}
  {\bibfnamefont {Vladan}\ \bibnamefont {Vuleti{\'{c}}}},\ }\bibfield  {title}
  {\enquote {\bibinfo {title} {{Near-Unitary Spin Squeezing in $^{171}$Yb}},}\
  }\href {\doibase 10.1103/PhysRevLett.122.223203} {\bibfield  {journal}
  {\bibinfo  {journal} {Phys. Rev. Lett.}\ }\textbf {\bibinfo {volume} {122}},\
  \bibinfo {pages} {223203} (\bibinfo {year} {2019})}\BibitemShut {NoStop}%
\bibitem [{\citenamefont {Kaubruegger}\ \emph {et~al.}(2019)\citenamefont
  {Kaubruegger}, \citenamefont {Silvi}, \citenamefont {Kokail}, \citenamefont
  {van Bijnen}, \citenamefont {Rey}, \citenamefont {Ye}, \citenamefont
  {Kaufman},\ and\ \citenamefont {Zoller}}]{Kaubruegger2019}%
  \BibitemOpen
  \bibfield  {author} {\bibinfo {author} {\bibfnamefont {Raphael}\ \bibnamefont
  {Kaubruegger}}, \bibinfo {author} {\bibfnamefont {Pietro}\ \bibnamefont
  {Silvi}}, \bibinfo {author} {\bibfnamefont {Christian}\ \bibnamefont
  {Kokail}}, \bibinfo {author} {\bibfnamefont {Rick}\ \bibnamefont {van
  Bijnen}}, \bibinfo {author} {\bibfnamefont {Ana~Maria}\ \bibnamefont {Rey}},
  \bibinfo {author} {\bibfnamefont {Jun}\ \bibnamefont {Ye}}, \bibinfo {author}
  {\bibfnamefont {Adam~M.}\ \bibnamefont {Kaufman}}, \ and\ \bibinfo {author}
  {\bibfnamefont {Peter}\ \bibnamefont {Zoller}},\ }\bibfield  {title}
  {\enquote {\bibinfo {title} {{Variational spin-squeezing algorithms on
  programmable quantum sensors}},}\ }\href {http://arxiv.org/abs/1908.08343}
  {\bibfield  {journal} {\bibinfo  {journal} {arXiv:1908.08343}\ } (\bibinfo
  {year} {2019})}\BibitemShut {NoStop}%
\bibitem [{\citenamefont {Koczor}\ \emph {et~al.}(2019)\citenamefont {Koczor},
  \citenamefont {Endo}, \citenamefont {Jones}, \citenamefont {Matsuzaki},\ and\
  \citenamefont {Benjamin}}]{Koczor2019}%
  \BibitemOpen
  \bibfield  {author} {\bibinfo {author} {\bibfnamefont {B{\'{a}}lint}\
  \bibnamefont {Koczor}}, \bibinfo {author} {\bibfnamefont {Suguru}\
  \bibnamefont {Endo}}, \bibinfo {author} {\bibfnamefont {Tyson}\ \bibnamefont
  {Jones}}, \bibinfo {author} {\bibfnamefont {Yuichiro}\ \bibnamefont
  {Matsuzaki}}, \ and\ \bibinfo {author} {\bibfnamefont {Simon~C.}\
  \bibnamefont {Benjamin}},\ }\bibfield  {title} {\enquote {\bibinfo {title}
  {{Variational-State Quantum Metrology}},}\ }\href
  {http://arxiv.org/abs/1908.08904} {\bibfield  {journal} {\bibinfo  {journal}
  {arXiv:1908.08904}\ } (\bibinfo {year} {2019})}\BibitemShut {NoStop}%
\bibitem [{\citenamefont {K{\'{o}}m{\'{a}}r}\ \emph {et~al.}(2014)\citenamefont
  {K{\'{o}}m{\'{a}}r}, \citenamefont {Kessler}, \citenamefont {Bishof},
  \citenamefont {Jiang}, \citenamefont {S{\o}rensen}, \citenamefont {Ye},\ and\
  \citenamefont {Lukin}}]{Komar2014}%
  \BibitemOpen
  \bibfield  {author} {\bibinfo {author} {\bibfnamefont {P.}~\bibnamefont
  {K{\'{o}}m{\'{a}}r}}, \bibinfo {author} {\bibfnamefont {E.~M.}\ \bibnamefont
  {Kessler}}, \bibinfo {author} {\bibfnamefont {M.}~\bibnamefont {Bishof}},
  \bibinfo {author} {\bibfnamefont {L.}~\bibnamefont {Jiang}}, \bibinfo
  {author} {\bibfnamefont {A.~S.}\ \bibnamefont {S{\o}rensen}}, \bibinfo
  {author} {\bibfnamefont {J.}~\bibnamefont {Ye}}, \ and\ \bibinfo {author}
  {\bibfnamefont {M.~D.}\ \bibnamefont {Lukin}},\ }\bibfield  {title} {\enquote
  {\bibinfo {title} {{A quantum network of clocks}},}\ }\href {\doibase
  10.1038/nphys3000} {\bibfield  {journal} {\bibinfo  {journal} {Nat. Phys.}\
  }\textbf {\bibinfo {volume} {10}},\ \bibinfo {pages} {582--587} (\bibinfo
  {year} {2014})}\BibitemShut {NoStop}%
\bibitem [{\citenamefont {Daley}\ \emph {et~al.}(2008)\citenamefont {Daley},
  \citenamefont {Boyd}, \citenamefont {Ye},\ and\ \citenamefont
  {Zoller}}]{Daley2008}%
  \BibitemOpen
  \bibfield  {author} {\bibinfo {author} {\bibfnamefont {Andrew~J.}\
  \bibnamefont {Daley}}, \bibinfo {author} {\bibfnamefont {Martin~M.}\
  \bibnamefont {Boyd}}, \bibinfo {author} {\bibfnamefont {Jun}\ \bibnamefont
  {Ye}}, \ and\ \bibinfo {author} {\bibfnamefont {Peter}\ \bibnamefont
  {Zoller}},\ }\bibfield  {title} {\enquote {\bibinfo {title} {{Quantum
  Computing with Alkaline-Earth-Metal Atoms}},}\ }\href {\doibase
  10.1103/PhysRevLett.101.170504} {\bibfield  {journal} {\bibinfo  {journal}
  {Phys. Rev. Lett.}\ }\textbf {\bibinfo {volume} {101}},\ \bibinfo {pages}
  {170504} (\bibinfo {year} {2008})}\BibitemShut {NoStop}%
\bibitem [{\citenamefont {Pagano}\ \emph {et~al.}(2019)\citenamefont {Pagano},
  \citenamefont {Scazza},\ and\ \citenamefont {FossFeig}}]{Pagano2018}%
  \BibitemOpen
  \bibfield  {author} {\bibinfo {author} {\bibfnamefont {Guido}\ \bibnamefont
  {Pagano}}, \bibinfo {author} {\bibfnamefont {Francesco}\ \bibnamefont
  {Scazza}}, \ and\ \bibinfo {author} {\bibfnamefont {Michael}\ \bibnamefont
  {Foss-Feig}},\ }\bibfield  {title} {\enquote {\bibinfo {title} {{Fast and
  Scalable Quantum Information Processing with Two-Electron Atoms in Optical
  Tweezer Arrays}},}\ }\href {\doibase 10.1002/qute.201800067} {\bibfield
  {journal} {\bibinfo  {journal} {Advanced Quantum Technologies}\ }\textbf
  {\bibinfo {volume} {2}},\ \bibinfo {pages} {1800067} (\bibinfo {year}
  {2019})}\BibitemShut {NoStop}%
\bibitem [{\citenamefont {Covey}\ \emph
  {et~al.}(2019{\natexlab{a}})\citenamefont {Covey}, \citenamefont {Sipahigil},
  \citenamefont {Szoke}, \citenamefont {Sinclair}, \citenamefont {Endres},\
  and\ \citenamefont {Painter}}]{Covey2019b}%
  \BibitemOpen
  \bibfield  {author} {\bibinfo {author} {\bibfnamefont {Jacob~P.}\
  \bibnamefont {Covey}}, \bibinfo {author} {\bibfnamefont {Alp}\ \bibnamefont
  {Sipahigil}}, \bibinfo {author} {\bibfnamefont {Szilard}\ \bibnamefont
  {Szoke}}, \bibinfo {author} {\bibfnamefont {Neil}\ \bibnamefont {Sinclair}},
  \bibinfo {author} {\bibfnamefont {Manuel}\ \bibnamefont {Endres}}, \ and\
  \bibinfo {author} {\bibfnamefont {Oskar}\ \bibnamefont {Painter}},\
  }\bibfield  {title} {\enquote {\bibinfo {title} {{Telecom-Band Quantum Optics
  with Ytterbium Atoms and Silicon Nanophotonics}},}\ }\href {\doibase
  10.1103/PhysRevApplied.11.034044} {\bibfield  {journal} {\bibinfo  {journal}
  {Phys. Rev. Applied}\ }\textbf {\bibinfo {volume} {11}},\ \bibinfo {pages}
  {034044} (\bibinfo {year} {2019}{\natexlab{a}})}\BibitemShut {NoStop}%
\bibitem [{\citenamefont {Huntemann}\ \emph {et~al.}(2012)\citenamefont
  {Huntemann}, \citenamefont {Okhapkin}, \citenamefont {Lipphardt},
  \citenamefont {Weyers}, \citenamefont {Tamm},\ and\ \citenamefont
  {Peik}}]{Huntemann2012}%
  \BibitemOpen
  \bibfield  {author} {\bibinfo {author} {\bibfnamefont {N.}~\bibnamefont
  {Huntemann}}, \bibinfo {author} {\bibfnamefont {M.}~\bibnamefont {Okhapkin}},
  \bibinfo {author} {\bibfnamefont {B.}~\bibnamefont {Lipphardt}}, \bibinfo
  {author} {\bibfnamefont {S.}~\bibnamefont {Weyers}}, \bibinfo {author}
  {\bibfnamefont {Chr.}\ \bibnamefont {Tamm}}, \ and\ \bibinfo {author}
  {\bibfnamefont {E.}~\bibnamefont {Peik}},\ }\bibfield  {title} {\enquote
  {\bibinfo {title} {{High-Accuracy Optical Clock Based on the Octupole
  Transition in $^{171}$Yb$^+$}},}\ }\href {\doibase
  10.1103/PhysRevLett.108.090801} {\bibfield  {journal} {\bibinfo  {journal}
  {Phys. Rev. Lett.}\ }\textbf {\bibinfo {volume} {108}},\ \bibinfo {pages}
  {090801} (\bibinfo {year} {2012})}\BibitemShut {NoStop}%
\bibitem [{\citenamefont {Tan}\ \emph {et~al.}(2019)\citenamefont {Tan},
  \citenamefont {Kaewuam}, \citenamefont {Arnold}, \citenamefont {Chanu},
  \citenamefont {Zhang}, \citenamefont {Safronova},\ and\ \citenamefont
  {Barrett}}]{Tan2019}%
  \BibitemOpen
  \bibfield  {author} {\bibinfo {author} {\bibfnamefont {T.~R.}\ \bibnamefont
  {Tan}}, \bibinfo {author} {\bibfnamefont {R.}~\bibnamefont {Kaewuam}},
  \bibinfo {author} {\bibfnamefont {K.~J.}\ \bibnamefont {Arnold}}, \bibinfo
  {author} {\bibfnamefont {S.~R.}\ \bibnamefont {Chanu}}, \bibinfo {author}
  {\bibfnamefont {Zhiqiang}\ \bibnamefont {Zhang}}, \bibinfo {author}
  {\bibfnamefont {M.~S.}\ \bibnamefont {Safronova}}, \ and\ \bibinfo {author}
  {\bibfnamefont {M.~D.}\ \bibnamefont {Barrett}},\ }\bibfield  {title}
  {\enquote {\bibinfo {title} {{Suppressing Inhomogeneous Broadening in a
  Lutetium Multi-ion Optical Clock}},}\ }\href {\doibase
  10.1103/PhysRevLett.123.063201} {\bibfield  {journal} {\bibinfo  {journal}
  {Phys. Rev. Lett.}\ }\textbf {\bibinfo {volume} {123}},\ \bibinfo {pages}
  {063201} (\bibinfo {year} {2019})}\BibitemShut {NoStop}%
\bibitem [{\citenamefont {Covey}\ \emph
  {et~al.}(2019{\natexlab{b}})\citenamefont {Covey}, \citenamefont {Madjarov},
  \citenamefont {Cooper},\ and\ \citenamefont {Endres}}]{Covey2019a}%
  \BibitemOpen
  \bibfield  {author} {\bibinfo {author} {\bibfnamefont {Jacob~P.}\
  \bibnamefont {Covey}}, \bibinfo {author} {\bibfnamefont {Ivaylo~S.}\
  \bibnamefont {Madjarov}}, \bibinfo {author} {\bibfnamefont {Alexandre}\
  \bibnamefont {Cooper}}, \ and\ \bibinfo {author} {\bibfnamefont {Manuel}\
  \bibnamefont {Endres}},\ }\bibfield  {title} {\enquote {\bibinfo {title}
  {{2000-Times Repeated Imaging of Strontium Atoms in Clock-Magic Tweezer
  Arrays}},}\ }\href {\doibase 10.1103/PhysRevLett.122.173201} {\bibfield
  {journal} {\bibinfo  {journal} {Phys. Rev. Lett.}\ }\textbf {\bibinfo
  {volume} {122}},\ \bibinfo {pages} {173201} (\bibinfo {year}
  {2019}{\natexlab{b}})}\BibitemShut {NoStop}%
\bibitem [{\citenamefont {Taichenachev}\ \emph {et~al.}(2006)\citenamefont
  {Taichenachev}, \citenamefont {Yudin}, \citenamefont {Oates}, \citenamefont
  {Hoyt}, \citenamefont {Barber},\ and\ \citenamefont
  {Hollberg}}]{Taichenachev2006a}%
  \BibitemOpen
  \bibfield  {author} {\bibinfo {author} {\bibfnamefont {A.}~\bibnamefont
  {Taichenachev}}, \bibinfo {author} {\bibfnamefont {V.}~\bibnamefont {Yudin}},
  \bibinfo {author} {\bibfnamefont {C.}~\bibnamefont {Oates}}, \bibinfo
  {author} {\bibfnamefont {C.}~\bibnamefont {Hoyt}}, \bibinfo {author}
  {\bibfnamefont {Z.}~\bibnamefont {Barber}}, \ and\ \bibinfo {author}
  {\bibfnamefont {L.}~\bibnamefont {Hollberg}},\ }\bibfield  {title} {\enquote
  {\bibinfo {title} {{Magnetic Field-Induced Spectroscopy of Forbidden Optical
  Transitions with Application to Lattice-Based Optical Atomic Clocks}},}\
  }\href {\doibase 10.1103/PhysRevLett.96.083001} {\bibfield  {journal}
  {\bibinfo  {journal} {Phys. Rev. Lett.}\ }\textbf {\bibinfo {volume} {96}},\
  \bibinfo {pages} {083001} (\bibinfo {year} {2006})}\BibitemShut {NoStop}%
\bibitem [{\citenamefont {Akatsuka}\ \emph {et~al.}(2010)\citenamefont
  {Akatsuka}, \citenamefont {Takamoto},\ and\ \citenamefont
  {Katori}}]{Akatsuka2010}%
  \BibitemOpen
  \bibfield  {author} {\bibinfo {author} {\bibfnamefont {Tomoya}\ \bibnamefont
  {Akatsuka}}, \bibinfo {author} {\bibfnamefont {Masao}\ \bibnamefont
  {Takamoto}}, \ and\ \bibinfo {author} {\bibfnamefont {Hidetoshi}\
  \bibnamefont {Katori}},\ }\bibfield  {title} {\enquote {\bibinfo {title}
  {{Three-dimensional optical lattice clock with bosonic $^{88}$Sr atoms}},}\
  }\href {\doibase 10.1103/PhysRevA.81.023402} {\bibfield  {journal} {\bibinfo
  {journal} {Phys. Rev. A}\ }\textbf {\bibinfo {volume} {81}},\ \bibinfo
  {pages} {023402} (\bibinfo {year} {2010})}\BibitemShut {NoStop}%
\bibitem [{\citenamefont {Nicholson}\ \emph {et~al.}(2015)\citenamefont
  {Nicholson}, \citenamefont {Campbell}, \citenamefont {Hutson}, \citenamefont
  {Marti}, \citenamefont {Bloom}, \citenamefont {McNally}, \citenamefont
  {Zhang}, \citenamefont {Barrett}, \citenamefont {Safronova}, \citenamefont
  {Strouse}, \citenamefont {Tew},\ and\ \citenamefont {Ye}}]{Nicholson2015}%
  \BibitemOpen
  \bibfield  {author} {\bibinfo {author} {\bibfnamefont {T.L.}\ \bibnamefont
  {Nicholson}}, \bibinfo {author} {\bibfnamefont {S.L.}\ \bibnamefont
  {Campbell}}, \bibinfo {author} {\bibfnamefont {R.B.}\ \bibnamefont {Hutson}},
  \bibinfo {author} {\bibfnamefont {G.E.}\ \bibnamefont {Marti}}, \bibinfo
  {author} {\bibfnamefont {B.J.}\ \bibnamefont {Bloom}}, \bibinfo {author}
  {\bibfnamefont {R.L.}\ \bibnamefont {McNally}}, \bibinfo {author}
  {\bibfnamefont {W.}~\bibnamefont {Zhang}}, \bibinfo {author} {\bibfnamefont
  {M.D.}\ \bibnamefont {Barrett}}, \bibinfo {author} {\bibfnamefont {M.S.}\
  \bibnamefont {Safronova}}, \bibinfo {author} {\bibfnamefont {G.F.}\
  \bibnamefont {Strouse}}, \bibinfo {author} {\bibfnamefont {W.L.}\
  \bibnamefont {Tew}}, \ and\ \bibinfo {author} {\bibfnamefont
  {J.}~\bibnamefont {Ye}},\ }\bibfield  {title} {\enquote {\bibinfo {title}
  {{Systematic evaluation of an atomic clock at 2 × 10-18 total
  uncertainty}},}\ }\href {\doibase 10.1038/ncomms7896} {\bibfield  {journal}
  {\bibinfo  {journal} {Nat. Commun.}\ }\textbf {\bibinfo {volume} {6}},\
  \bibinfo {pages} {6896} (\bibinfo {year} {2015})}\BibitemShut {NoStop}%
\bibitem [{\citenamefont {Al-Masoudi}\ \emph {et~al.}(2015)\citenamefont
  {Al-Masoudi}, \citenamefont {D{\"{o}}rscher}, \citenamefont {H{\"{a}}fner},
  \citenamefont {Sterr},\ and\ \citenamefont {Lisdat}}]{Al-Masoudi2015}%
  \BibitemOpen
  \bibfield  {author} {\bibinfo {author} {\bibfnamefont {Ali}\ \bibnamefont
  {Al-Masoudi}}, \bibinfo {author} {\bibfnamefont {S{\"{o}}ren}\ \bibnamefont
  {D{\"{o}}rscher}}, \bibinfo {author} {\bibfnamefont {Sebastian}\ \bibnamefont
  {H{\"{a}}fner}}, \bibinfo {author} {\bibfnamefont {Uwe}\ \bibnamefont
  {Sterr}}, \ and\ \bibinfo {author} {\bibfnamefont {Christian}\ \bibnamefont
  {Lisdat}},\ }\bibfield  {title} {\enquote {\bibinfo {title} {{Noise and
  instability of an optical lattice clock}},}\ }\href {\doibase
  10.1103/PhysRevA.92.063814} {\bibfield  {journal} {\bibinfo  {journal} {Phys.
  Rev. A}\ }\textbf {\bibinfo {volume} {92}},\ \bibinfo {pages} {063814}
  (\bibinfo {year} {2015})}\BibitemShut {NoStop}%
\bibitem [{\citenamefont {Brown}\ \emph {et~al.}(2017)\citenamefont {Brown},
  \citenamefont {Phillips}, \citenamefont {Beloy}, \citenamefont {McGrew},
  \citenamefont {Schioppo}, \citenamefont {Fasano}, \citenamefont {Milani},
  \citenamefont {Zhang}, \citenamefont {Hinkley}, \citenamefont {Leopardi},
  \citenamefont {Yoon}, \citenamefont {Nicolodi}, \citenamefont {Fortier},\
  and\ \citenamefont {Ludlow}}]{Brown2017a}%
  \BibitemOpen
  \bibfield  {author} {\bibinfo {author} {\bibfnamefont {R.~C.}\ \bibnamefont
  {Brown}}, \bibinfo {author} {\bibfnamefont {N.~B.}\ \bibnamefont {Phillips}},
  \bibinfo {author} {\bibfnamefont {K.}~\bibnamefont {Beloy}}, \bibinfo
  {author} {\bibfnamefont {W.~F.}\ \bibnamefont {McGrew}}, \bibinfo {author}
  {\bibfnamefont {M.}~\bibnamefont {Schioppo}}, \bibinfo {author}
  {\bibfnamefont {R.~J.}\ \bibnamefont {Fasano}}, \bibinfo {author}
  {\bibfnamefont {G.}~\bibnamefont {Milani}}, \bibinfo {author} {\bibfnamefont
  {X.}~\bibnamefont {Zhang}}, \bibinfo {author} {\bibfnamefont
  {N.}~\bibnamefont {Hinkley}}, \bibinfo {author} {\bibfnamefont
  {H.}~\bibnamefont {Leopardi}}, \bibinfo {author} {\bibfnamefont {T.~H.}\
  \bibnamefont {Yoon}}, \bibinfo {author} {\bibfnamefont {D.}~\bibnamefont
  {Nicolodi}}, \bibinfo {author} {\bibfnamefont {T.~M.}\ \bibnamefont
  {Fortier}}, \ and\ \bibinfo {author} {\bibfnamefont {A.~D.}\ \bibnamefont
  {Ludlow}},\ }\bibfield  {title} {\enquote {\bibinfo {title}
  {{Hyperpolarizability and Operational Magic Wavelength in an Optical Lattice
  Clock}},}\ }\href {\doibase 10.1103/PhysRevLett.119.253001} {\bibfield
  {journal} {\bibinfo  {journal} {Phys. Rev. Lett.}\ }\textbf {\bibinfo
  {volume} {119}},\ \bibinfo {pages} {253001} (\bibinfo {year}
  {2017})}\BibitemShut {NoStop}%
\bibitem [{\citenamefont {Origlia}\ \emph {et~al.}(2018)\citenamefont
  {Origlia}, \citenamefont {Pramod}, \citenamefont {Schiller}, \citenamefont
  {Singh}, \citenamefont {Bongs}, \citenamefont {Schwarz}, \citenamefont
  {Al-Masoudi}, \citenamefont {D{\"{o}}rscher}, \citenamefont {Herbers},
  \citenamefont {H{\"{a}}fner}, \citenamefont {Sterr},\ and\ \citenamefont
  {Lisdat}}]{Origlia2018}%
  \BibitemOpen
  \bibfield  {author} {\bibinfo {author} {\bibfnamefont {S.}~\bibnamefont
  {Origlia}}, \bibinfo {author} {\bibfnamefont {M.~S.}\ \bibnamefont {Pramod}},
  \bibinfo {author} {\bibfnamefont {S.}~\bibnamefont {Schiller}}, \bibinfo
  {author} {\bibfnamefont {Y.}~\bibnamefont {Singh}}, \bibinfo {author}
  {\bibfnamefont {K.}~\bibnamefont {Bongs}}, \bibinfo {author} {\bibfnamefont
  {R.}~\bibnamefont {Schwarz}}, \bibinfo {author} {\bibfnamefont
  {A.}~\bibnamefont {Al-Masoudi}}, \bibinfo {author} {\bibfnamefont
  {S.}~\bibnamefont {D{\"{o}}rscher}}, \bibinfo {author} {\bibfnamefont
  {S.}~\bibnamefont {Herbers}}, \bibinfo {author} {\bibfnamefont
  {S.}~\bibnamefont {H{\"{a}}fner}}, \bibinfo {author} {\bibfnamefont
  {U.}~\bibnamefont {Sterr}}, \ and\ \bibinfo {author} {\bibfnamefont {Ch.}\
  \bibnamefont {Lisdat}},\ }\bibfield  {title} {\enquote {\bibinfo {title}
  {{Towards an optical clock for space: Compact, high-performance optical
  lattice clock based on bosonic atoms}},}\ }\href {\doibase
  10.1103/PhysRevA.98.053443} {\bibfield  {journal} {\bibinfo  {journal} {Phys.
  Rev. A}\ }\textbf {\bibinfo {volume} {98}},\ \bibinfo {pages} {053443}
  (\bibinfo {year} {2018})}\BibitemShut {NoStop}%
\bibitem [{\citenamefont {Nemitz}\ \emph {et~al.}(2019)\citenamefont {Nemitz},
  \citenamefont {J{\o}rgensen}, \citenamefont {Yanagimoto}, \citenamefont
  {Bregolin},\ and\ \citenamefont {Katori}}]{Nemitz2019}%
  \BibitemOpen
  \bibfield  {author} {\bibinfo {author} {\bibfnamefont {Nils}\ \bibnamefont
  {Nemitz}}, \bibinfo {author} {\bibfnamefont {Asbj{\o}rn~Arvad}\ \bibnamefont
  {J{\o}rgensen}}, \bibinfo {author} {\bibfnamefont {Ryotatsu}\ \bibnamefont
  {Yanagimoto}}, \bibinfo {author} {\bibfnamefont {Filippo}\ \bibnamefont
  {Bregolin}}, \ and\ \bibinfo {author} {\bibfnamefont {Hidetoshi}\
  \bibnamefont {Katori}},\ }\bibfield  {title} {\enquote {\bibinfo {title}
  {{Modeling light shifts in optical lattice clocks}},}\ }\href {\doibase
  10.1103/PhysRevA.99.033424} {\bibfield  {journal} {\bibinfo  {journal} {Phys.
  Rev. A}\ }\textbf {\bibinfo {volume} {99}},\ \bibinfo {pages} {033424}
  (\bibinfo {year} {2019})}\BibitemShut {NoStop}%
\bibitem [{\citenamefont {Koller}\ \emph {et~al.}(2017)\citenamefont {Koller},
  \citenamefont {Grotti}, \citenamefont {Vogt}, \citenamefont {Al-Masoudi},
  \citenamefont {D{\"{o}}rscher}, \citenamefont {H{\"{a}}fner}, \citenamefont
  {Sterr},\ and\ \citenamefont {Lisdat}}]{Koller2017a}%
  \BibitemOpen
  \bibfield  {author} {\bibinfo {author} {\bibfnamefont {S.~B.}\ \bibnamefont
  {Koller}}, \bibinfo {author} {\bibfnamefont {J.}~\bibnamefont {Grotti}},
  \bibinfo {author} {\bibfnamefont {St.}\ \bibnamefont {Vogt}}, \bibinfo
  {author} {\bibfnamefont {A.}~\bibnamefont {Al-Masoudi}}, \bibinfo {author}
  {\bibfnamefont {S.}~\bibnamefont {D{\"{o}}rscher}}, \bibinfo {author}
  {\bibfnamefont {S.}~\bibnamefont {H{\"{a}}fner}}, \bibinfo {author}
  {\bibfnamefont {U.}~\bibnamefont {Sterr}}, \ and\ \bibinfo {author}
  {\bibfnamefont {Ch.}\ \bibnamefont {Lisdat}},\ }\bibfield  {title} {\enquote
  {\bibinfo {title} {{Transportable Optical Lattice Clock with
  $7\times10^{-17}$ Uncertainty}},}\ }\href {\doibase
  10.1103/PhysRevLett.118.073601} {\bibfield  {journal} {\bibinfo  {journal}
  {Phys. Rev. Lett.}\ }\textbf {\bibinfo {volume} {118}},\ \bibinfo {pages}
  {073601} (\bibinfo {year} {2017})}\BibitemShut {NoStop}%
\bibitem [{\citenamefont {Norcia}\ \emph {et~al.}(2019)\citenamefont {Norcia},
  \citenamefont {Young}, \citenamefont {Eckner}, \citenamefont {Oelker},
  \citenamefont {Ye},\ and\ \citenamefont {Kaufman}}]{Norcia2019}%
  \BibitemOpen
  \bibfield  {author} {\bibinfo {author} {\bibfnamefont {Matthew~A.}\
  \bibnamefont {Norcia}}, \bibinfo {author} {\bibfnamefont {Aaron~W.}\
  \bibnamefont {Young}}, \bibinfo {author} {\bibfnamefont {William~J.}\
  \bibnamefont {Eckner}}, \bibinfo {author} {\bibfnamefont {Eric}\ \bibnamefont
  {Oelker}}, \bibinfo {author} {\bibfnamefont {Jun}\ \bibnamefont {Ye}}, \ and\
  \bibinfo {author} {\bibfnamefont {Adam~M.}\ \bibnamefont {Kaufman}},\
  }\bibfield  {title} {\enquote {\bibinfo {title} {Seconds-scale coherence on
  an optical clock transition in a tweezer array},}\ }\href {\doibase
  10.1126/science.aay0644} {\bibfield  {journal} {\bibinfo  {journal}
  {Science}\ }\textbf {\bibinfo {volume} {366}},\ \bibinfo {pages} {93--97}
  (\bibinfo {year} {2019})}\BibitemShut {NoStop}%
\bibitem [{\citenamefont {Dick}(1987)}]{Dick1987}%
  \BibitemOpen
  \bibfield  {author} {\bibinfo {author} {\bibfnamefont {G.~John}\ \bibnamefont
  {Dick}},\ }\bibfield  {title} {\enquote {\bibinfo {title} {{Local Oscillator
  Induced Instabilities in Trapped Ion Frequency Standards}},}\ }\href@noop {}
  {\bibfield  {journal} {\bibinfo  {journal} {Proceedings of the 19th Annual
  Precise Time and Time Interval Systems and Applications}\ ,\ \bibinfo {pages}
  {133 -- 147}} (\bibinfo {year} {1987})}\BibitemShut {NoStop}%
\bibitem [{\citenamefont {Campbell}\ \emph {et~al.}(2017)\citenamefont
  {Campbell}, \citenamefont {Hutson}, \citenamefont {Marti}, \citenamefont
  {Goban}, \citenamefont {{Darkwah Oppong}}, \citenamefont {McNally},
  \citenamefont {Sonderhouse}, \citenamefont {Robinson}, \citenamefont {Zhang},
  \citenamefont {Bloom},\ and\ \citenamefont {Ye}}]{Campbell2017}%
  \BibitemOpen
  \bibfield  {author} {\bibinfo {author} {\bibfnamefont {S.~L.}\ \bibnamefont
  {Campbell}}, \bibinfo {author} {\bibfnamefont {R.~B.}\ \bibnamefont
  {Hutson}}, \bibinfo {author} {\bibfnamefont {G.~E.}\ \bibnamefont {Marti}},
  \bibinfo {author} {\bibfnamefont {A.}~\bibnamefont {Goban}}, \bibinfo
  {author} {\bibfnamefont {N.}~\bibnamefont {{Darkwah Oppong}}}, \bibinfo
  {author} {\bibfnamefont {R.~L.}\ \bibnamefont {McNally}}, \bibinfo {author}
  {\bibfnamefont {L.}~\bibnamefont {Sonderhouse}}, \bibinfo {author}
  {\bibfnamefont {J.~M.}\ \bibnamefont {Robinson}}, \bibinfo {author}
  {\bibfnamefont {W.}~\bibnamefont {Zhang}}, \bibinfo {author} {\bibfnamefont
  {B.~J.}\ \bibnamefont {Bloom}}, \ and\ \bibinfo {author} {\bibfnamefont
  {J.}~\bibnamefont {Ye}},\ }\bibfield  {title} {\enquote {\bibinfo {title} {{A
  Fermi-degenerate three-dimensional optical lattice clock}},}\ }\href
  {\doibase 10.1126/science.aam5538} {\bibfield  {journal} {\bibinfo  {journal}
  {Science}\ }\textbf {\bibinfo {volume} {358}},\ \bibinfo {pages} {90--94}
  (\bibinfo {year} {2017})}\BibitemShut {NoStop}%
\bibitem [{\citenamefont {Schioppo}\ \emph {et~al.}(2017)\citenamefont
  {Schioppo}, \citenamefont {Brown}, \citenamefont {McGrew}, \citenamefont
  {Hinkley}, \citenamefont {Fasano}, \citenamefont {Beloy}, \citenamefont
  {Yoon}, \citenamefont {Milani}, \citenamefont {Nicolodi}, \citenamefont
  {Sherman}, \citenamefont {Phillips}, \citenamefont {Oates},\ and\
  \citenamefont {Ludlow}}]{Schioppo2017}%
  \BibitemOpen
  \bibfield  {author} {\bibinfo {author} {\bibfnamefont {M.}~\bibnamefont
  {Schioppo}}, \bibinfo {author} {\bibfnamefont {R.~C.}\ \bibnamefont {Brown}},
  \bibinfo {author} {\bibfnamefont {W.~F.}\ \bibnamefont {McGrew}}, \bibinfo
  {author} {\bibfnamefont {N.}~\bibnamefont {Hinkley}}, \bibinfo {author}
  {\bibfnamefont {R.~J.}\ \bibnamefont {Fasano}}, \bibinfo {author}
  {\bibfnamefont {K.}~\bibnamefont {Beloy}}, \bibinfo {author} {\bibfnamefont
  {T.~H.}\ \bibnamefont {Yoon}}, \bibinfo {author} {\bibfnamefont
  {G.}~\bibnamefont {Milani}}, \bibinfo {author} {\bibfnamefont
  {D.}~\bibnamefont {Nicolodi}}, \bibinfo {author} {\bibfnamefont {J.~A.}\
  \bibnamefont {Sherman}}, \bibinfo {author} {\bibfnamefont {N.~B.}\
  \bibnamefont {Phillips}}, \bibinfo {author} {\bibfnamefont {C.~W.}\
  \bibnamefont {Oates}}, \ and\ \bibinfo {author} {\bibfnamefont {A.~D.}\
  \bibnamefont {Ludlow}},\ }\bibfield  {title} {\enquote {\bibinfo {title}
  {{Ultrastable optical clock with two cold-atom ensembles}},}\ }\href
  {\doibase 10.1038/nphoton.2016.231} {\bibfield  {journal} {\bibinfo
  {journal} {Nat. Photonics}\ }\textbf {\bibinfo {volume} {11}},\ \bibinfo
  {pages} {48--52} (\bibinfo {year} {2017})}\BibitemShut {NoStop}%
\bibitem [{\citenamefont {Swallows}\ \emph {et~al.}(2011)\citenamefont
  {Swallows}, \citenamefont {Bishof}, \citenamefont {Lin}, \citenamefont
  {Blatt}, \citenamefont {Martin}, \citenamefont {Rey},\ and\ \citenamefont
  {Ye}}]{Swallows2011}%
  \BibitemOpen
  \bibfield  {author} {\bibinfo {author} {\bibfnamefont {M.~D.}\ \bibnamefont
  {Swallows}}, \bibinfo {author} {\bibfnamefont {M.}~\bibnamefont {Bishof}},
  \bibinfo {author} {\bibfnamefont {Y.}~\bibnamefont {Lin}}, \bibinfo {author}
  {\bibfnamefont {S.}~\bibnamefont {Blatt}}, \bibinfo {author} {\bibfnamefont
  {M.~J.}\ \bibnamefont {Martin}}, \bibinfo {author} {\bibfnamefont {A.~M.}\
  \bibnamefont {Rey}}, \ and\ \bibinfo {author} {\bibfnamefont
  {J.}~\bibnamefont {Ye}},\ }\bibfield  {title} {\enquote {\bibinfo {title}
  {{Suppression of Collisional Shifts in a Strongly Interacting Lattice
  Clock}},}\ }\href {\doibase 10.1126/science.1196442} {\bibfield  {journal}
  {\bibinfo  {journal} {Science}\ }\textbf {\bibinfo {volume} {331}},\ \bibinfo
  {pages} {1043--1046} (\bibinfo {year} {2011})}\BibitemShut {NoStop}%
\bibitem [{\citenamefont {Chang}\ \emph {et~al.}(2004)\citenamefont {Chang},
  \citenamefont {Ye},\ and\ \citenamefont {Lukin}}]{Chang2004}%
  \BibitemOpen
  \bibfield  {author} {\bibinfo {author} {\bibfnamefont {D.~E.}\ \bibnamefont
  {Chang}}, \bibinfo {author} {\bibfnamefont {Jun}\ \bibnamefont {Ye}}, \ and\
  \bibinfo {author} {\bibfnamefont {M.~D.}\ \bibnamefont {Lukin}},\ }\bibfield
  {title} {\enquote {\bibinfo {title} {{Controlling dipole-dipole frequency
  shifts in a lattice-based optical atomic clock}},}\ }\href {\doibase
  10.1103/PhysRevA.69.023810} {\bibfield  {journal} {\bibinfo  {journal} {Phys.
  Rev. A}\ }\textbf {\bibinfo {volume} {69}},\ \bibinfo {pages} {023810}
  (\bibinfo {year} {2004})}\BibitemShut {NoStop}%
\bibitem [{\citenamefont {Hutson}\ \emph {et~al.}(2019)\citenamefont {Hutson},
  \citenamefont {Goban}, \citenamefont {Marti}, \citenamefont {Sonderhouse},
  \citenamefont {Sanner},\ and\ \citenamefont {Ye}}]{Hutson2019}%
  \BibitemOpen
  \bibfield  {author} {\bibinfo {author} {\bibfnamefont {Ross~B.}\ \bibnamefont
  {Hutson}}, \bibinfo {author} {\bibfnamefont {Akihisa}\ \bibnamefont {Goban}},
  \bibinfo {author} {\bibfnamefont {G.~Edward}\ \bibnamefont {Marti}}, \bibinfo
  {author} {\bibfnamefont {Lindsay}\ \bibnamefont {Sonderhouse}}, \bibinfo
  {author} {\bibfnamefont {Christian}\ \bibnamefont {Sanner}}, \ and\ \bibinfo
  {author} {\bibfnamefont {Jun}\ \bibnamefont {Ye}},\ }\bibfield  {title}
  {\enquote {\bibinfo {title} {{Engineering Quantum States of Matter for Atomic
  Clocks in Shallow Optical Lattices}},}\ }\href
  {http://arxiv.org/abs/1903.02498} {\bibfield  {journal} {\bibinfo  {journal}
  {arXiv:1903.02498}\ } (\bibinfo {year} {2019})}\BibitemShut {NoStop}%
\bibitem [{\citenamefont {Cooper}\ \emph {et~al.}(2018)\citenamefont {Cooper},
  \citenamefont {Covey}, \citenamefont {Madjarov}, \citenamefont {Porsev},
  \citenamefont {Safronova},\ and\ \citenamefont {Endres}}]{Cooper2018a}%
  \BibitemOpen
  \bibfield  {author} {\bibinfo {author} {\bibfnamefont {Alexandre}\
  \bibnamefont {Cooper}}, \bibinfo {author} {\bibfnamefont {Jacob~P.}\
  \bibnamefont {Covey}}, \bibinfo {author} {\bibfnamefont {Ivaylo~S.}\
  \bibnamefont {Madjarov}}, \bibinfo {author} {\bibfnamefont {Sergey~G.}\
  \bibnamefont {Porsev}}, \bibinfo {author} {\bibfnamefont {Marianna~S.}\
  \bibnamefont {Safronova}}, \ and\ \bibinfo {author} {\bibfnamefont {Manuel}\
  \bibnamefont {Endres}},\ }\bibfield  {title} {\enquote {\bibinfo {title}
  {{Alkaline-Earth Atoms in Optical Tweezers}},}\ }\href {\doibase
  10.1103/PhysRevX.8.041055} {\bibfield  {journal} {\bibinfo  {journal} {Phys.
  Rev. X}\ }\textbf {\bibinfo {volume} {8}},\ \bibinfo {pages} {041055}
  (\bibinfo {year} {2018})}\BibitemShut {NoStop}%
\bibitem [{\citenamefont {Nogrette}\ \emph {et~al.}(2014)\citenamefont
  {Nogrette}, \citenamefont {Labuhn}, \citenamefont {Ravets}, \citenamefont
  {Barredo}, \citenamefont {B{\'{e}}guin}, \citenamefont {Vernier},
  \citenamefont {Lahaye},\ and\ \citenamefont {Browaeys}}]{Nogrette2014}%
  \BibitemOpen
  \bibfield  {author} {\bibinfo {author} {\bibfnamefont {F.}~\bibnamefont
  {Nogrette}}, \bibinfo {author} {\bibfnamefont {H.}~\bibnamefont {Labuhn}},
  \bibinfo {author} {\bibfnamefont {S.}~\bibnamefont {Ravets}}, \bibinfo
  {author} {\bibfnamefont {D.}~\bibnamefont {Barredo}}, \bibinfo {author}
  {\bibfnamefont {L.}~\bibnamefont {B{\'{e}}guin}}, \bibinfo {author}
  {\bibfnamefont {A.}~\bibnamefont {Vernier}}, \bibinfo {author} {\bibfnamefont
  {T.}~\bibnamefont {Lahaye}}, \ and\ \bibinfo {author} {\bibfnamefont
  {A.}~\bibnamefont {Browaeys}},\ }\bibfield  {title} {\enquote {\bibinfo
  {title} {{Single-Atom Trapping in Holographic 2D Arrays of Microtraps with
  Arbitrary Geometries}},}\ }\href {\doibase 10.1103/PhysRevX.4.021034}
  {\bibfield  {journal} {\bibinfo  {journal} {Phys. Rev. X}\ }\textbf {\bibinfo
  {volume} {4}},\ \bibinfo {pages} {021034} (\bibinfo {year}
  {2014})}\BibitemShut {NoStop}%
\bibitem [{\citenamefont {Riehle}(2003)}]{Riehle2003}%
  \BibitemOpen
  \bibfield  {author} {\bibinfo {author} {\bibfnamefont {Fritz}\ \bibnamefont
  {Riehle}},\ }\href {\doibase 10.1002/3527605991} {\emph {\bibinfo {title}
  {{Frequency Standards}}}}\ (\bibinfo  {publisher} {Wiley},\ \bibinfo {year}
  {2003})\BibitemShut {NoStop}%
\bibitem [{\citenamefont {Norcia}\ \emph {et~al.}(2018)\citenamefont {Norcia},
  \citenamefont {Young},\ and\ \citenamefont {Kaufman}}]{Norcia2018b}%
  \BibitemOpen
  \bibfield  {author} {\bibinfo {author} {\bibfnamefont {M.~A.}\ \bibnamefont
  {Norcia}}, \bibinfo {author} {\bibfnamefont {A.~W.}\ \bibnamefont {Young}}, \
  and\ \bibinfo {author} {\bibfnamefont {A.~M.}\ \bibnamefont {Kaufman}},\
  }\bibfield  {title} {\enquote {\bibinfo {title} {{Microscopic Control and
  Detection of Ultracold Strontium in Optical-Tweezer Arrays}},}\ }\href
  {\doibase 10.1103/PhysRevX.8.041054} {\bibfield  {journal} {\bibinfo
  {journal} {Phys. Rev. X}\ }\textbf {\bibinfo {volume} {8}},\ \bibinfo {pages}
  {041054} (\bibinfo {year} {2018})}\BibitemShut {NoStop}%
\bibitem [{\citenamefont {Norcia}(2019)}]{Norcia2019b}%
  \BibitemOpen
  \bibfield  {author} {\bibinfo {author} {\bibfnamefont {Matthew~A.}\
  \bibnamefont {Norcia}},\ }\bibfield  {title} {\enquote {\bibinfo {title}
  {{Coupling atoms to cavities with narrow linewidth optical transitions:
  Applications to frequency metrology}},}\ }\href
  {http://arxiv.org/abs/1908.11442} {\bibfield  {journal} {\bibinfo  {journal}
  {arXiv:1908.11442}\ } (\bibinfo {year} {2019})}\BibitemShut {NoStop}%
\bibitem [{\citenamefont {Wineland}\ and\ \citenamefont
  {Itano}(1979)}]{Wineland1979}%
  \BibitemOpen
  \bibfield  {author} {\bibinfo {author} {\bibfnamefont {D.~J.}\ \bibnamefont
  {Wineland}}\ and\ \bibinfo {author} {\bibfnamefont {Wayne~M.}\ \bibnamefont
  {Itano}},\ }\bibfield  {title} {\enquote {\bibinfo {title} {{Laser cooling of
  atoms}},}\ }\href {\doibase 10.1103/PhysRevA.20.1521} {\bibfield  {journal}
  {\bibinfo  {journal} {Phys. Rev. A}\ }\textbf {\bibinfo {volume} {20}},\
  \bibinfo {pages} {1521--1540} (\bibinfo {year} {1979})}\BibitemShut {NoStop}%
\bibitem [{\citenamefont {de~L{\'{e}}s{\'{e}}leuc}\ \emph
  {et~al.}(2018)\citenamefont {de~L{\'{e}}s{\'{e}}leuc}, \citenamefont
  {Barredo}, \citenamefont {Lienhard}, \citenamefont {Browaeys},\ and\
  \citenamefont {Lahaye}}]{deLeseleuc2018}%
  \BibitemOpen
  \bibfield  {author} {\bibinfo {author} {\bibfnamefont {Sylvain}\ \bibnamefont
  {de~L{\'{e}}s{\'{e}}leuc}}, \bibinfo {author} {\bibfnamefont {Daniel}\
  \bibnamefont {Barredo}}, \bibinfo {author} {\bibfnamefont {Vincent}\
  \bibnamefont {Lienhard}}, \bibinfo {author} {\bibfnamefont {Antoine}\
  \bibnamefont {Browaeys}}, \ and\ \bibinfo {author} {\bibfnamefont {Thierry}\
  \bibnamefont {Lahaye}},\ }\bibfield  {title} {\enquote {\bibinfo {title}
  {{Analysis of imperfections in the coherent optical excitation of single
  atoms to Rydberg states}},}\ }\href {\doibase 10.1103/PhysRevA.97.053803}
  {\bibfield  {journal} {\bibinfo  {journal} {Phys. Rev. A}\ }\textbf {\bibinfo
  {volume} {97}},\ \bibinfo {pages} {053803} (\bibinfo {year}
  {2018})}\BibitemShut {NoStop}%
\bibitem [{\citenamefont {Numata}\ \emph {et~al.}(2004)\citenamefont {Numata},
  \citenamefont {Kemery},\ and\ \citenamefont {Camp}}]{Numata2004}%
  \BibitemOpen
  \bibfield  {author} {\bibinfo {author} {\bibfnamefont {Kenji}\ \bibnamefont
  {Numata}}, \bibinfo {author} {\bibfnamefont {Amy}\ \bibnamefont {Kemery}}, \
  and\ \bibinfo {author} {\bibfnamefont {Jordan}\ \bibnamefont {Camp}},\
  }\bibfield  {title} {\enquote {\bibinfo {title} {{Thermal-Noise Limit in the
  Frequency Stabilization of Lasers with Rigid Cavities}},}\ }\href {\doibase
  10.1103/PhysRevLett.93.250602} {\bibfield  {journal} {\bibinfo  {journal}
  {Phys. Rev. Lett.}\ }\textbf {\bibinfo {volume} {93}},\ \bibinfo {pages}
  {250602} (\bibinfo {year} {2004})}\BibitemShut {NoStop}%
\bibitem [{\citenamefont {Jiang}\ \emph {et~al.}(2011)\citenamefont {Jiang},
  \citenamefont {Ludlow}, \citenamefont {Lemke}, \citenamefont {Fox},
  \citenamefont {Sherman}, \citenamefont {Ma},\ and\ \citenamefont
  {Oates}}]{Jiang2011}%
  \BibitemOpen
  \bibfield  {author} {\bibinfo {author} {\bibfnamefont {Y.~Y.}\ \bibnamefont
  {Jiang}}, \bibinfo {author} {\bibfnamefont {A.~D.}\ \bibnamefont {Ludlow}},
  \bibinfo {author} {\bibfnamefont {N.~D.}\ \bibnamefont {Lemke}}, \bibinfo
  {author} {\bibfnamefont {R.~W.}\ \bibnamefont {Fox}}, \bibinfo {author}
  {\bibfnamefont {J.~A.}\ \bibnamefont {Sherman}}, \bibinfo {author}
  {\bibfnamefont {L.-S.}\ \bibnamefont {Ma}}, \ and\ \bibinfo {author}
  {\bibfnamefont {C.~W.}\ \bibnamefont {Oates}},\ }\bibfield  {title} {\enquote
  {\bibinfo {title} {{Making optical atomic clocks more stable with
  $10^{-16}$-level laser stabilization}},}\ }\href {\doibase
  10.1038/nphoton.2010.313} {\bibfield  {journal} {\bibinfo  {journal} {Nat.
  Photonics}\ }\textbf {\bibinfo {volume} {5}},\ \bibinfo {pages} {158--161}
  (\bibinfo {year} {2011})}\BibitemShut {NoStop}%
\bibitem [{\citenamefont {Webster}\ and\ \citenamefont
  {Gill}(2011)}]{Webster2011}%
  \BibitemOpen
  \bibfield  {author} {\bibinfo {author} {\bibfnamefont {Stephen}\ \bibnamefont
  {Webster}}\ and\ \bibinfo {author} {\bibfnamefont {Patrick}\ \bibnamefont
  {Gill}},\ }\bibfield  {title} {\enquote {\bibinfo {title} {{Force-insensitive
  optical cavity}},}\ }\href {\doibase 10.1364/OL.36.003572} {\bibfield
  {journal} {\bibinfo  {journal} {Opt. Lett.}\ }\textbf {\bibinfo {volume}
  {36}},\ \bibinfo {pages} {3572} (\bibinfo {year} {2011})}\BibitemShut
  {NoStop}%
\bibitem [{\citenamefont {Barber}\ \emph {et~al.}(2006)\citenamefont {Barber},
  \citenamefont {Hoyt}, \citenamefont {Oates}, \citenamefont {Hollberg},
  \citenamefont {Taichenachev},\ and\ \citenamefont {Yudin}}]{Barber2006}%
  \BibitemOpen
  \bibfield  {author} {\bibinfo {author} {\bibfnamefont {Z.}~\bibnamefont
  {Barber}}, \bibinfo {author} {\bibfnamefont {C.}~\bibnamefont {Hoyt}},
  \bibinfo {author} {\bibfnamefont {C.}~\bibnamefont {Oates}}, \bibinfo
  {author} {\bibfnamefont {L.}~\bibnamefont {Hollberg}}, \bibinfo {author}
  {\bibfnamefont {A.}~\bibnamefont {Taichenachev}}, \ and\ \bibinfo {author}
  {\bibfnamefont {V.}~\bibnamefont {Yudin}},\ }\bibfield  {title} {\enquote
  {\bibinfo {title} {{Direct Excitation of the Forbidden Clock Transition in
  Neutral $^{174}$Yb Atoms Confined to an Optical Lattice}},}\ }\href {\doibase
  10.1103/PhysRevLett.96.083002} {\bibfield  {journal} {\bibinfo  {journal}
  {Phys. Rev. Lett.}\ }\textbf {\bibinfo {volume} {96}},\ \bibinfo {pages}
  {083002} (\bibinfo {year} {2006})}\BibitemShut {NoStop}%
\bibitem [{\citenamefont {Han}\ \emph {et~al.}(2018)\citenamefont {Han},
  \citenamefont {Zhou}, \citenamefont {Zhang}, \citenamefont {Gao},
  \citenamefont {Xu}, \citenamefont {Li}, \citenamefont {Zhang},\ and\
  \citenamefont {Xu}}]{Han2018}%
  \BibitemOpen
  \bibfield  {author} {\bibinfo {author} {\bibfnamefont {Chengyin}\
  \bibnamefont {Han}}, \bibinfo {author} {\bibfnamefont {Min}\ \bibnamefont
  {Zhou}}, \bibinfo {author} {\bibfnamefont {Xiaohang}\ \bibnamefont {Zhang}},
  \bibinfo {author} {\bibfnamefont {Qi}~\bibnamefont {Gao}}, \bibinfo {author}
  {\bibfnamefont {Yilin}\ \bibnamefont {Xu}}, \bibinfo {author} {\bibfnamefont
  {Shangyan}\ \bibnamefont {Li}}, \bibinfo {author} {\bibfnamefont {Shuang}\
  \bibnamefont {Zhang}}, \ and\ \bibinfo {author} {\bibfnamefont {Xinye}\
  \bibnamefont {Xu}},\ }\bibfield  {title} {\enquote {\bibinfo {title}
  {{Carrier thermometry of cold ytterbium atoms in an optical lattice
  clock}},}\ }\href {\doibase 10.1038/s41598-018-26367-8} {\bibfield  {journal}
  {\bibinfo  {journal} {Scientific Reports}\ }\textbf {\bibinfo {volume} {8}},\
  \bibinfo {pages} {7927} (\bibinfo {year} {2018})}\BibitemShut {NoStop}%
\bibitem [{\citenamefont {Katori}\ \emph {et~al.}(2015)\citenamefont {Katori},
  \citenamefont {Ovsiannikov}, \citenamefont {Marmo},\ and\ \citenamefont
  {Palchikov}}]{Katori2015}%
  \BibitemOpen
  \bibfield  {author} {\bibinfo {author} {\bibfnamefont {Hidetoshi}\
  \bibnamefont {Katori}}, \bibinfo {author} {\bibfnamefont {V.~D.}\
  \bibnamefont {Ovsiannikov}}, \bibinfo {author} {\bibfnamefont {S.~I.}\
  \bibnamefont {Marmo}}, \ and\ \bibinfo {author} {\bibfnamefont {V.~G.}\
  \bibnamefont {Palchikov}},\ }\bibfield  {title} {\enquote {\bibinfo {title}
  {{Strategies for reducing the light shift in atomic clocks}},}\ }\href
  {\doibase 10.1103/PhysRevA.91.052503} {\bibfield  {journal} {\bibinfo
  {journal} {Phys. Rev. A}\ }\textbf {\bibinfo {volume} {91}},\ \bibinfo
  {pages} {052503} (\bibinfo {year} {2015})}\BibitemShut {NoStop}%
\bibitem [{\citenamefont {{Le Targat}}\ \emph {et~al.}(2013)\citenamefont {{Le
  Targat}}, \citenamefont {Lorini}, \citenamefont {{Le Coq}}, \citenamefont
  {Zawada}, \citenamefont {Gu{\'{e}}na}, \citenamefont {Abgrall}, \citenamefont
  {Gurov}, \citenamefont {Rosenbusch}, \citenamefont {Rovera}, \citenamefont
  {Nag{\'{o}}rny}, \citenamefont {Gartman}, \citenamefont {Westergaard},
  \citenamefont {Tobar}, \citenamefont {Lours}, \citenamefont {Santarelli},
  \citenamefont {Clairon}, \citenamefont {Bize}, \citenamefont {Laurent},
  \citenamefont {Lemonde},\ and\ \citenamefont {Lodewyck}}]{LeTargat2013}%
  \BibitemOpen
  \bibfield  {author} {\bibinfo {author} {\bibfnamefont {R.}~\bibnamefont {{Le
  Targat}}}, \bibinfo {author} {\bibfnamefont {L.}~\bibnamefont {Lorini}},
  \bibinfo {author} {\bibfnamefont {Y.}~\bibnamefont {{Le Coq}}}, \bibinfo
  {author} {\bibfnamefont {M.}~\bibnamefont {Zawada}}, \bibinfo {author}
  {\bibfnamefont {J.}~\bibnamefont {Gu{\'{e}}na}}, \bibinfo {author}
  {\bibfnamefont {M.}~\bibnamefont {Abgrall}}, \bibinfo {author} {\bibfnamefont
  {M.}~\bibnamefont {Gurov}}, \bibinfo {author} {\bibfnamefont
  {P.}~\bibnamefont {Rosenbusch}}, \bibinfo {author} {\bibfnamefont {D.~G.}\
  \bibnamefont {Rovera}}, \bibinfo {author} {\bibfnamefont {B.}~\bibnamefont
  {Nag{\'{o}}rny}}, \bibinfo {author} {\bibfnamefont {R.}~\bibnamefont
  {Gartman}}, \bibinfo {author} {\bibfnamefont {P.~G.}\ \bibnamefont
  {Westergaard}}, \bibinfo {author} {\bibfnamefont {M.~E.}\ \bibnamefont
  {Tobar}}, \bibinfo {author} {\bibfnamefont {M.}~\bibnamefont {Lours}},
  \bibinfo {author} {\bibfnamefont {G.}~\bibnamefont {Santarelli}}, \bibinfo
  {author} {\bibfnamefont {A.}~\bibnamefont {Clairon}}, \bibinfo {author}
  {\bibfnamefont {S.}~\bibnamefont {Bize}}, \bibinfo {author} {\bibfnamefont
  {P.}~\bibnamefont {Laurent}}, \bibinfo {author} {\bibfnamefont
  {P.}~\bibnamefont {Lemonde}}, \ and\ \bibinfo {author} {\bibfnamefont
  {J.}~\bibnamefont {Lodewyck}},\ }\bibfield  {title} {\enquote {\bibinfo
  {title} {{Experimental realization of an optical second with strontium
  lattice clocks}},}\ }\href {\doibase 10.1038/ncomms3109} {\bibfield
  {journal} {\bibinfo  {journal} {Nat. Commun.}\ }\textbf {\bibinfo {volume}
  {4}},\ \bibinfo {pages} {2109} (\bibinfo {year} {2013})}\BibitemShut
  {NoStop}%
\bibitem [{\citenamefont {Middelmann}\ \emph {et~al.}(2012)\citenamefont
  {Middelmann}, \citenamefont {Falke}, \citenamefont {Lisdat},\ and\
  \citenamefont {Sterr}}]{Middelmann2012}%
  \BibitemOpen
  \bibfield  {author} {\bibinfo {author} {\bibfnamefont {Thomas}\ \bibnamefont
  {Middelmann}}, \bibinfo {author} {\bibfnamefont {Stephan}\ \bibnamefont
  {Falke}}, \bibinfo {author} {\bibfnamefont {Christian}\ \bibnamefont
  {Lisdat}}, \ and\ \bibinfo {author} {\bibfnamefont {Uwe}\ \bibnamefont
  {Sterr}},\ }\bibfield  {title} {\enquote {\bibinfo {title} {{High Accuracy
  Correction of Blackbody Radiation Shift in an Optical Lattice Clock}},}\
  }\href {\doibase 10.1103/PhysRevLett.109.263004} {\bibfield  {journal}
  {\bibinfo  {journal} {Phys. Rev. Lett.}\ }\textbf {\bibinfo {volume} {109}},\
  \bibinfo {pages} {263004} (\bibinfo {year} {2012})}\BibitemShut {NoStop}%
\bibitem [{\citenamefont {Westergaard}\ \emph {et~al.}(2011)\citenamefont
  {Westergaard}, \citenamefont {Lodewyck}, \citenamefont {Lorini},
  \citenamefont {Lecallier}, \citenamefont {Burt}, \citenamefont {Zawada},
  \citenamefont {Millo},\ and\ \citenamefont {Lemonde}}]{Westergaard2011}%
  \BibitemOpen
  \bibfield  {author} {\bibinfo {author} {\bibfnamefont {P.~G.}\ \bibnamefont
  {Westergaard}}, \bibinfo {author} {\bibfnamefont {J.}~\bibnamefont
  {Lodewyck}}, \bibinfo {author} {\bibfnamefont {L.}~\bibnamefont {Lorini}},
  \bibinfo {author} {\bibfnamefont {A.}~\bibnamefont {Lecallier}}, \bibinfo
  {author} {\bibfnamefont {E.~A.}\ \bibnamefont {Burt}}, \bibinfo {author}
  {\bibfnamefont {M.}~\bibnamefont {Zawada}}, \bibinfo {author} {\bibfnamefont
  {J.}~\bibnamefont {Millo}}, \ and\ \bibinfo {author} {\bibfnamefont
  {P.}~\bibnamefont {Lemonde}},\ }\bibfield  {title} {\enquote {\bibinfo
  {title} {{Lattice-Induced Frequency Shifts in Sr Optical Lattice Clocks at
  the $10^{-17}$ Level}},}\ }\href {\doibase 10.1103/PhysRevLett.106.210801}
  {\bibfield  {journal} {\bibinfo  {journal} {Phys. Rev. Lett.}\ }\textbf
  {\bibinfo {volume} {106}},\ \bibinfo {pages} {210801} (\bibinfo {year}
  {2011})}\BibitemShut {NoStop}%
\end{thebibliography}
\end{document}